%% file: multpap.tex
\documentstyle[12pt,epsfig,rotating,cite]{article}
\newcommand{\avgn}{\langle n\rangle}
\newcommand{\anch}{\langle n_{ch}\rangle}
\newcommand{\nbar}{\overline{n}}
\newcommand{\chnd}{$\chi^2/NDF$}
\newcommand{\chndf}[2]{\multicolumn{4}{c}{{#1}/{#2}}}
\newcommand{\dgr}{^\circ}
\newcommand{\ra}{\rightarrow}


\newcommand{\aver}[1]{\left<#1\right>}

\newcommand{\pbinv}{\mbox{${\rm ~pb}^{-1}$}}
\newcommand{\etas}{\eta^\ast}

\newcommand{\pp}{\mbox{$p\bar{p}$\ }}

\newcommand{\epp}{\mbox{$e^+p$\ }}
\newcommand{\gsp}{\mbox{$\gamma^\ast p$\ }}

\newcommand{\gevsq}{${\rm GeV}^2$}

\newcommand{\MD}{multiplicity distribution}
\newcommand{\MDS}{multiplicity distributions}

\newcommand{\AVN}{\aver{n_{ch}}}
\renewcommand{\AVN}{\aver{n}}
\newcommand{\CHDF}{\chi^2/\mbox{\small NDF}}

\begin{document}
\bibliographystyle{unsrt}

%=======================================================================

\begin{titlepage}

\noindent{\tt DESY 96-160 \hfill ISSN 0418-9833}\\
\tt August 1996\hfill\break

\vspace*{4.cm}

\begin{center}
\begin{Large}
\bf{Charged Particle Multiplicities \\in Deep Inelastic Scattering at
HERA\\}
\vspace*{2.cm}
H1 Collaboration
\end{Large}
\end{center}

\vspace*{2cm}

\begin{abstract} 
\noindent Using the H1 detector at HERA, charged particle multiplicity
distributions in deep inelastic \epp scattering have been measured over
a large kinematical region.  The evolution with $W$ and $Q^2$ of the
multiplicity distribution and of the multiplicity moments in
pseudorapidity domains of varying size is studied in the current
fragmentation region of the hadronic centre-of-mass frame.  The results
are compared with data from fixed target lepton-nucleon interactions,
$e^+e^-$ annihilations and hadron-hadron collisions as well as with
expectations from QCD based parton models.  Fits to the Negative
Binomial and Lognormal distributions are presented.
\end{abstract}

\end{titlepage}

%=======================================================================

\vfill
\cleardoublepage

{\tolerance = 10000
\input h1auts.tex

\vspace*{0.5cm}
\input h1inst.tex
}
%=======================================================================

\newpage

\input body.tex

%% file: h1auts.tex
%-- H1AUTS  Author list by names
%--Status: 08/05/96
\noindent S.~Aid$^{13}$,                   %HAM2-PD      8/93        Aid
 M.~Anderson$^{23}$,              %MANC-ST  10/95           Anderson
 V.~Andreev$^{26}$,               %LPI -PD                  Andreev
 B.~Andrieu$^{29}$,               %ECPL-PD                  Andrieu
 R.-D.~Appuhn$^{11}$,             %DESY-LEFT  10/95         Appuhn
%C.~Arndt$^{11}$,                 %DESY-STn  1/96           Arndt
 A.~Babaev$^{25}$,                %ITEP-PD                  Babaev
 J.~B\"ahr$^{36}$,                %ZEUT-PD                  Baehr
 J.~B\'an$^{18}$,                 %KOSI-PD                  Banj
 Y.~Ban$^{28}$,                   %ORSa-ST                  Bany
 P.~Baranov$^{26}$
% ...

%% file: h1inst.tex
%     H1 Institutes as appearing on publications
\noindent $ ^1$ I. Physikalisches Institut der RWTH, Aachen, Germany$^ a$ \\
 $ ^2$ III. Physikalisches Institut der RWTH, Aachen, Germany$^ a$ \\
%$ ^3$ Institut f\"ur Physik, Humboldt-Universit\"at,
%              Berlin, Germany$^ a$ \\
 $ ^3$ School of Physics and Space Research, University of Birmingham,
                             Birmingham, UK$^ b$\\
 $ ^4$ Inter-University Institute for High Energies ULB-VUB, Brussels;
   Universitaire Instelling Antwerpen, Wilrijk; Belgium$^ c$ \\
 $ ^5$ Rutherford Appleton Laboratory, Chilton, Didcot, UK$^ b$ \\
 $ ^6$ Institute for Nuclear Physics, Cracow, Poland$^ d$  \\
 $ ^7$ Physics Department and IIRPA,
         University of California, Davis, California, USA$^ e$ \\
 $ ^8$ Institut f\"ur Physik, Universit\"at Dortmund, Dortmund,
                                                  Germany$^ a$\\
 $ ^{9}$ CEA, DSM/DAPNIA, CE-Saclay, Gif-sur-Yvette, France \\
 $ ^{10}$ Department of Physics and Astronomy, University of Glasgow,
                                      Glasgow, UK$^ b$ \\
 $ ^{11}$ DESY, Hamburg, Germany$^a$ \\
 $ ^{12}$ I. Institut f\"ur Experimentalphysik, Universit\"at Hamburg,
                                     Hamburg, Germany$^ a$  \\
 $ ^{13}$ II. Institut f\"ur Experimentalphysik, Universit\"at Hamburg,
                                     Hamburg, Germany$^ a$  \\
 $ ^{14}$ Max-Planck-Institut f\"ur Kernphysik,
                                     Heidelberg, Germany$^ a$ \\
 $ ^{15}$ Physikalisches Institut, Universit\"at Heidelberg,
                                     Heidelberg, Germany$^ a$ \\
 $ ^{16}$ Institut f\"ur Hochenergiephysik, Universit\"at Heidelberg,
                                     Heidelberg, Germany$^ a$ \\
 $ ^{17}$ Institut f\"ur Reine und Angewandte Kernphysik, Universit\"at
                                   Kiel, Kiel, Germany$^ a$\\
 $ ^{18}$ Institute of Experimental Physics, Slovak Academy of
                Sciences, Ko\v{s}ice, Slovak Republic$^ f$\\
 $ ^{19}$ School of Physics and Chemistry, University of Lancaster,
                              Lancaster, UK$^ b$ \\
 $ ^{20}$ Department of Physics, University of Liverpool,
                                              Liverpool, UK$^ b$ \\
 $ ^{21}$ Queen Mary and Westfield College, London, UK$^ b$ \\
 $ ^{22}$ Physics Department, University of Lund,
                                               Lund, Sweden$^ g$ \\
 $ ^{23}$ Physics Department, University of Manchester,
                                          Manchester, UK$^ b$\\
 $ ^{24}$ CPPM, Universit\'{e} d'Aix-Marseille II,
                          IN2P3-CNRS, Marseille, France\\
 $ ^{25}$ Institute for Theoretical and Experimental Physics,
                                                 Moscow, Russia \\
 $ ^{26}$ Lebedev Physical Institute, Moscow, Russia$^ f$ \\
 $ ^{27}$ Max-Planck-Institut f\"ur Physik,
                                            M\"unchen, Germany$^ a$\\
 $ ^{28}$ LAL, Universit\'{e} de Paris-Sud, IN2P3-CNRS,
                            Orsay, France\\
 $ ^{29}$ LPNHE, Ecole Polytechnique, IN2P3-CNRS,
                             Palaiseau, France \\
 $ ^{30}$ LPNHE, Universit\'{e}s Paris VI and VII, IN2P3-CNRS,
                              Paris, France \\
 $ ^{31}$ Institute of  Physics, Czech Academy of
                    Sciences, Praha, Czech Republic$^{ f,h}$ \\
 $ ^{32}$ Nuclear Center, Charles University,
                    Praha, Czech Republic$^{ f,h}$ \\
%$ ^{33}$ INFN Roma and Dipartimento di Fisica,
%              Universita "La Sapienza", Roma, Italy   \\
 $ ^{33}$ INFN Roma~1 and Dipartimento di Fisica,
               Universit\`a Roma~3, Roma, Italy   \\
 $ ^{34}$ Paul Scherrer Institut, Villigen, Switzerland \\
 $ ^{35}$ Fachbereich Physik, Bergische Universit\"at Gesamthochschule
               Wuppertal, Wuppertal, Germany$^ a$ \\
 $ ^{36}$ DESY, Institut f\"ur Hochenergiephysik,
                              Zeuthen, Germany$^ a$\\
 $ ^{37}$ Institut f\"ur Teilchenphysik,
          ETH, Z\"urich, Switzerland$^ i$\\
 $ ^{38}$ Physik-Institut der Universit\"at Z\"urich,
                              Z\"urich, Switzerland$^ i$\\
\smallskip
 $ ^{39}$ Visitor from Yerevan Phys. Inst., Armenia\\
 $ ^{40}$ On leave from LBL, Berkeley, USA \\
 $ ^{41}$ Institut f\"ur Physik, Humboldt-Universit\"at,
               Berlin, Germany$^ a$ \\
 
\smallskip
\noindent $ ^{\dagger}$ Deceased\\
 
\bigskip
\noindent $ ^a$ Supported by the Bundesministerium f\"ur Bildung, Wissenschaft,
        Forschung und Technologie, FRG,
        under contract numbers 6AC17P, 6AC47P, 6DO57I, 6HH17P, 6HH27I,
        6HD17I, 6HD27I, 6KI17P, 6MP17I, and 6WT87P \\
 $ ^b$ Supported by the UK Particle Physics and Astronomy Research
       Council, and formerly by the UK Science and Engineering Research
       Council \\
 $ ^c$ Supported by FNRS-NFWO, IISN-IIKW \\
 $ ^d$ Supported by the Polish State Committee for Scientific Research,
       grant \hfil\\ nos. 115/E-743/SPUB/P03/109/95 and 2~P03B~244~08p01,
       and Stiftung f\"ur Deutsch-Polnische Zusammenarbeit,
       project no.506/92 \\
 $ ^e$ Supported in part by USDOE grant DE~F603~91ER40674\\
 $ ^f$ Supported by the Deutsche Forschungsgemeinschaft\\
 $ ^g$ Supported by the Swedish Natural Science Research Council\\
 $ ^h$ Supported by GA \v{C}R, grant no. 202/93/2423,
       GA AV \v{C}R, grant no. 19095 and GA UK, grant no. 342\\
 $ ^i$ Supported by the Swiss National Science Foundation\\

%% file: body.tex
\section{Introduction}

The multiplicity distribution of hadrons produced in high energy
interactions is one of the basic measures characterising multiparticle
final states.  The fluctuation pattern of the number of particles
produced in a given domain of phase space reveals the nature of the
correlations among the hadrons and is, therefore, sensitive to the
dynamics of the process. The multiplicity distribution of charged
hadrons has been measured for the full variety of collision processes
from $e^+e^-$ annihilations to nucleus-nucleus collisions and the available
data cover a wide energy range~\cite{mul:exp:rev,schmitz:rev}. Numerous
theoretical studies have been devoted to the
subject~\cite{carr:shih,dremin:rev,ugoccioni}, starting with early
investigations by Heisenberg and Fermi~\cite{heisenberg:fermi}.

Whereas the total event multiplicity remains of considerable importance,
interest has shifted with time towards studies of the multiplicity
distribution in subdomains of phase space. In these restricted domains,
global conservation constraints are minimised and dynamical correlation
effects better revealed~\cite{bialas:hayot,review:93b}.

In this paper we present results on the multiplicity distribution of
charged hadrons produced in the current fragmentation region of
deep inelastic scattering (DIS) \epp collisions. The analysis is based
on data accumulated by the H1 detector at HERA in 1994, corresponding to
an integrated luminosity of 1.3\pbinv.

The multiplicity distribution and its moments, fully corrected for
detector effects, are studied in the virtual-boson proton (\gsp)
rest system, the hadronic centre-of-mass frame.  The measurements
are obtained in subdomains of pseudorapidity
space\footnote{Pseudorapidity is defined as $\etas = -\ln
\tan \frac{\theta}{2}$, with $\theta$ the angle between the hadron
momentum and the direction of the virtual photon in the \gsp rest
system. The current hemisphere is defined as the region of
positive~$\etas$.}, both as a function of $W$, the total hadronic
centre-of-mass energy, and $Q^2$, the negative four-momentum transfer
squared in the DIS process.  Results from deep inelastic \epp
interactions from the HERA collider on the charged multiplicity
distribution~\cite{zeus:md} and on the mean multiplicity in the current
region of the Breit-frame~\cite{h1:thompson} have already been
published.

Here, H1 results are compared with those obtained in DIS fixed target
experiments at much lower energy, with hadroproduction in $e^+e^-$
annihilation reactions, with data from hadron-hadron collisions as
well as with the expectations of QCD based parton shower models and other
more phenomenological approaches.

%=======================================================================

\section{Definitions and phenomenology\label{sec:feno:qcd}}

%=======================================================================

\subsection{Correlations\label{sec:corr}}

The set of probabilities $P_n$ for various numbers of charged hadrons
($n$) to be produced in a given region of phase space is known as the
multiplicity distribution. The continued interest in multiplicity
distributions rests on  the observation that fluctuations in
``counting statistics'' are a direct measure of the strength of
correlations among the objects being counted.  In high energy physics,
this was first exploited by Mueller~\cite{Mue71} in the
formulation of the concept of short range order.  Since then,
correlations have been extensively used as probes of the
interaction and hadronisation dynamics~\cite{review:93b}.

%a fundamental property of multiperipheral and multi-Regge dynamics, 
%and of Feynman's parton model~\cite{feynman:69}.  
 
The connection between the multiplicity distribution and particle
correlations is made explicit in the relation between the 
factorial moment of order $q$, $\tilde{R}_q$, of the
multiplicity distribution in a phase space domain (say, rapidity $\Delta$)
 and the $q$-particle inclusive momentum
density $\rho_q(1,2,\ldots,q)$ which for identical particles
takes the form~\cite{Mue71}:
\begin{equation}
\tilde{R}_q = \sum_0^\infty n(n-1)\ldots(n-q+1)\,P_n
= \int_{\Delta}\ldots\int_{\Delta}\,\rho_q(1,2,\ldots,q)\,d{y_1}\ldots
d{y_q}.  
\label{f:facmom} 
\end{equation} 
%
%\noindent In the above formula, we consider identical particles produced
%in a given phase space domain (say, rapidity interval) $\Delta$.
%
The functions $\rho_q$ contain, besides genuine dynamical or kinematical
correlations, contributions from ``random coincidences'' in the region
$\Delta$.  The latter are eliminated by considering the (``connected'')
correlation functions $\kappa_q(1,2,\ldots,q)$, familiar from
statistical physics.  Integrated over a domain of size $\Delta$ they 
define the factorial
cumulants (or Mueller moments) $\tilde{K}_q$ of the multiplicity
distribution\cite{Mue71}.
%
%The distribution $P_n$ is conveniently represented by the generating
%function
%\begin{equation}
%G(u)=\sum_{n=0}^\infty P_n\,(1+u)^n.\label{f:gener}
%\end{equation}
%From the series expansions~\cite{kendall}:
%\begin{eqnarray}
%G(u)&=&1+\aver{n}u+ \sum_{q=2}^\infty \frac{u^q}{q!}\,
%\tilde{R}_q,\label{f:fms}\\
%\ln G(u)&=&\aver{n}u+ \sum_{q=2}^\infty \frac{u^q}{q!}\,\tilde{K}_q;\label{f:cum}
%\end{eqnarray}
%one derives the relations:
%
For the normalised quantities $R_q=\tilde{R}_q/\aver{n}^q$ and
$K_q=\tilde{K_q}/\aver{n}^q$ the following relations hold:
\begin{eqnarray} 
K_2&=&R_2-1, \\ K_3&=&R_3-3R_2+2;
\end{eqnarray}
where $\aver{n}$ is the mean multiplicity in $\Delta$.
%=======================================================================

\subsection{Scaling}

Koba, Nielsen and Olesen (KNO)~\cite{KNO} have studied the limit of the
multiplicity distribution in full  phase space 
for $n\rightarrow\infty$ and
$\avgn\rightarrow\infty$ with $z=n/\avgn$ fixed. In this case one
obtains the KNO form
\begin{equation}
\aver{n}P_n\simeq\Psi(z).\label{f:knoform}
\end{equation}
 The function $\Psi(z)$ was shown to become asymptotically independent of
the total energy if Feynman scaling~\cite{feynman:69} is satisfied.  A
proper mathematical reformulation of multiplicity scaling for discrete
distributions, valid also at finite energies, was later given by
Golokhvastov~\cite{golov} and is known as KNO-G scaling.

Exact KNO scaling implies that, besides $\Psi(z)$,  the factorial
moments $R_q$ as well as the moments
$C_q\equiv\aver{n^q}/{\avgn}^q$ are energy independent. 
%Data on these quantities will be presented in Sect.~\ref{sec:kno}.

Scaling, either in KNO or in  KNO-G form, is  experimentally
well-established~\cite{ochs:md:90,wrob:logn} for the full phase space and
single hemisphere multiplicity distributions  in $e^+e^-$ annihilations, in
DIS lepton-hadron interactions and in hadron-hadron
collisions~\cite{szwed:wrochna:85}, except at the highest  SPS collider
energies~\cite{Carl87:2,ua5:md:2:9}.  This is remarkable 
since Feynman scaling is strongly violated in all these
processes. However, already in 1970, Polyakov~\cite{scale:cascade}
derived the KNO scaling law for $e^+e^-$ hadronic final states within a broad
class of field theories. This pre-QCD model is based on a conformal
invariance principle which can be reformulated in terms of a
scale-invariant stochastic branching process. An energy independent
coupling constant is assumed and the mean multiplicity rises as a power
of the energy.
The Polyakov derivation, and subsequent work~\cite{scale:cascade},
demonstrates that KNO scaling should hold  at least approximately
in any model based on similar principles.
%including the approximately
%scale-invariant QCD parton shower model~\cite{basetto,konishi}.

%=======================================================================

\subsection{Parametric models\label{sec:param:models}}

%Many discrete distributions of phenomenological interest belong to the
%class of Poisson transforms~\cite{CarShih}. They are represented as a
%continuous superposition of Poisson distributions:

%\begin{equation}
%P_n=\int_0^\infty dz\,\Psi(z)\, \frac{\left(z\aver{n}\right)^n}{n!}\,
%e^{-z\aver{n}}\,\,\,\,n=0,1,\ldots.\label{poiss:tr}
%\end{equation}

%with $\aver{n}$ the mean multiplicity.  The function $\Psi(z)$ satisfies
%the normalisation conditions
%$\int_0^\infty\,\Psi(z)dz=\int_0^\infty\,z\Psi(z)dz=1$.  It follows that
%$R_q=\int_0^\infty z^q\,\Psi(z)\,dz$:  the normalised factorial moments of
%the discrete distribution $P_n$ are equal to the ordinary moments of the
%continuous distribution $\Psi(z)$. 

A much studied distribution is the Negative Binomial Distribution (NBD)
defined as
\begin{equation}
P_n(k,\nbar)=
\frac{k(k+1)\cdots(k+n-1)}{n!}\,
\left(\frac{\nbar}{\nbar+k} \right)^n \,
\left(\frac{k}{\nbar+k}\right)^k,\label{nbd:1}
\end{equation}
\noindent with parameters $k$ (or $1/k$) and $\nbar$.  The
average $\avgn$ and the dispersion $D$ of the NBD are
related to the two parameters by
\begin{equation}
\avgn=\nbar\,\,\,\,\mbox{;}\,\,\,\,\frac{D^2}{\avgn^2}
=\frac{1}{\nbar} +\frac{1}{k}.\label{nbd:2}
\end{equation}
\noindent For $1/k\rightarrow0$  (\ref{nbd:1}) reduces to a Poissonian.
%The factorial cumulants  $K_q =(q-1)!\,(1/k)^{q-1}$ are independent of
$\nbar$. 
The parameter $1/k$ is equal to the integrated
second order correlation function $K_2$  in the studied phase space domain.

%The density function $\Psi(z)$ in (\ref{poiss:tr}) 
%is a Gamma distribution with parameter $k$~\cite{carr:shih}.

Many phenomenological models for hadroproduction 
predict multiplicity distributions of a Negative Binomial form.
They are reviewed
in~\cite{carr:shih,vanhove:1}.  In QCD, the NBD is obtained as a solution in
Leading-Log Approximation for the gluon multiplicity distribution in a
quark jet~\cite{Giovan79}.  
Within the framework of the Modified Leading-Log Approximation (MLLA) and Local
Parton-Hadron Duality (LPHD)~\cite{lphd}, the lower order factorial moments 
behave approximately as those of the
NBD~\cite{malaza:webber,dipole:model,doksh:93}.

It is well-established experimentally in a large variety of collision
processes that multiplicity distributions, in full phase space as well
as in restricted phase space domains, are approximately of Negative
Binomial form.  However, deviations are observed in
high statistics $e^+e^-$ experiments for final states with hard
jets~\cite{delphi:md,aleph:rap:mul} and in hadron-hadron
interactions~\cite{ua5:md:2:9}.  In fixed target DIS lepton-hadron and
lepton-nucleus collisions the NBD adequately describes
the multiplicity distributions in the full phase space,  in limited
(pseudo-) rapidity intervals, as well as in the full current and target
hemispheres~\cite{schmitz:rev,e665:mul}.

A multiplicity distribution exhibiting explicit KNO-G scaling 
has recently been derived by assuming a scale-invariant 
bi-variate {\em multiplicative}
branching mechanism as the basis of multihadron
production~\cite{wrob:logn}. Application of the central limit theorem
leads to a scaling function $\Psi$ of Lognormal form.  In this model the
multiplicity distribution $P_n$ is related to a continuous probability
density $f(\tilde{n})$ and defined as $P_n=\int_n^{n+1}f(\tilde{n})d\tilde{n}$,
where $f(\tilde{n})$ is the Lognormal distribution (LND).  The mean
continuous multiplicity $\aver{\tilde{n}}$ is approximately
given by
$\aver{\tilde{n}}=\aver{n}+0.5$.  Following~\cite{wrob:logn} one finds
\begin{equation}
P_n=\int_{ n/\aver{\tilde{n}}}^{(n+1)/\aver{\tilde{n}}}
\frac{N}{\sqrt{2\pi}\sigma}\,\frac{1}{\tilde{z}+c}
\, \exp\left( -\frac{[\ln(\tilde{z}+c)-\mu]^2}%
{2\sigma^2}   \right)\,d\tilde{z}.\label{f:lnd}
\end{equation}
\noindent The integrand in (\ref{f:lnd}) defines the Lognormal
scaling function in KNO-G form. It depends on two parameters. Here,
$\tilde{z}=\tilde{n}/\aver{\tilde{n}}$ is the scaled continuous
multiplicity;  $N$, $\sigma$, $\mu$ and $c$ are parameters of which only
two are independent due to normalisation conditions.  In fits to data,
correlations between the parameters are reduced if $d$ and $c$ are used
as free parameters. In terms of $\sigma^2$ and $\mu$ these are given by

\begin{equation}
\sigma=\sqrt{\ln\left[ \left( \frac{d}{1+c}\right)^2+1\right] }\,\,\,\mbox{and}\,\,\,
\mu=\ln(c+1)-\sigma^2/2.\label{lnd:sig:mu}
\end{equation}

The parameter $d$ is equal to the dispersion of the scaling function. 
The expression for $P_n$ also depends on $\aver{\tilde{n}}$ or
$\avgn$. The latter fit parameter is denoted by $m$ in
Table~\ref{tab:fitpar}. Exact scaling implies that $d$ and $c$ are
energy independent.  In addition, the generalised dispersions
$D_q=\left[\aver{ (n-\avgn)^q} \right]^{1/q} $ ($D\equiv D_2$) 
satisfy a generalised Wr\'oblewski relation~\cite{wrob:law}.  Finally,
the mean continuous multiplicity is found to grow as a power of the
energy.  Comparisons of the Lognormal distribution with data from $e^+e^-$
annihilation, $\nu(\overline{\nu})p$ and \pp collisions can be found
in~\cite{aleph:rap:mul,wrob:logn,opal:mul,aleph:mul,jones:92}.

%In the simple quark-parton model (QPM), the
%virtual boson interacts with a quark (antiquark) of fractional momentum
%$x$ in the proton and transfers a four-momentum $q$.
%%
%The  nature of generic multiparticle final states in DIS is, at small $x$,
%expected to be a hybrid of $e^+e^-$ final state properties and
%generic hadron-hadron final states~\cite{bj:73}.
%In  QPM, and in any reference frame where the
%exchanged boson ($\gamma^\star$) and the target proton are
%collinear, the struck quark and the target remnant are separated
%by a rapidity  interval
% of length $\sim\ln{W^2}$. This  rapidity interval is filled with
%colour radiation and, eventually, with hadrons.
%The general expectations, in the absence of extra hard QCD jets, are
%that the rapidity region near the 
%virtual photon---the current region---behaves as in $e^+e^-$  annihilation, 
%while the rest has properties typical of  hadron-hadron collisions~\cite{bj:vietri}.
%The dividing line is a distance proportional to $\ln{Q^2}$
%from the leading virtual photon fragments
%and a distance $\sim\ln{(1/x-1)}$
%from the leading proton fragments. The sum is proportional to $\ln{W^2}$,
%as required.
%The hadrons produced in the rapidity region between current and proton 
%remnant are expected to be most sensitive to the QCD evolution 
%undergone by  the proton.

%=======================================================================

\section{Experimental procedure}

%=======================================================================

\subsection{The experiment}

The experiment was carried out with the H1 detector~\cite{h1:detector}
at the HERA storage ring at DESY. The data were collected during the
1994 running period when positrons, with an incident energy of 27.5 GeV,
collided with protons with an energy of 820 GeV.  The following briefly
describes the detector components most relevant to this analysis.

The  energy of the scattered positrons is
measured with a liquid argon (LAr) calorimeter and a ``backward''
electromagnetic lead-scintillator calorimeter (BEMC). The LAr
calorimeter~\cite{h1:calo} extends over the polar angular range $4\dgr <
\theta < 153\dgr$ with full azimuthal coverage, where $\theta$ is
defined with respect to the proton beam direction (positive $z$ axis).
It consists of an electromagnetic section with lead absorbers and a
hadronic section with steel absorbers.

%The hadronic energy scale and resolution have been
%studied using transverse momentum balance between the hadronic final state
%and the scattered positron in DIS events and are known to an accuracy of
%$5\%$ and $10\%$, respectively.

The BEMC covers the polar angular range $151\dgr < \theta <177\dgr$. 
A principal task of the BEMC is to trigger on and measure scattered
positrons in DIS processes with $Q^2$ values ranging from 5 to
100~\gevsq. The BEMC energy scale for positrons is known to an
accuracy of $1\%$.

The central tracking detectors are a hybrid of inner and outer
cylindrical jet chambers (CJC1 and CJC2), $z$-drift chambers and
proportional chambers. The latter provide a fast signal and allow H1 to
trigger on tracks which originate from the $z$ range expected for \epp
collisions.

The jet chambers, mounted concentrically around the beam line, provide
particle charge and momentum measurements 
from track curvature and cover the angular
interval $15\dgr < \theta < 165\dgr$.  Up to 56 space points can be
measured for non-curling tracks.

%The radial coordinates are derived from drift 
%time measurements, the $z$ coordinates from charge division.

The calorimeters and central trackers are surrounded by a
superconducting solenoid providing a uniform magnetic field of 1.15~T
parallel to the beam axis in the tracking region.

A backward proportional chamber (BPC), situated immediately in front of
the BEMC and with an angular acceptance of $155.5\dgr < \theta <
174.5\dgr$, serves to measure the impact point of the scattered
positron and to confirm that the particle entering the BEMC is charged.
Using information from the BPC, the BEMC and the reconstructed event
vertex, the polar angle of the scattered positron  can be determined to better
than 1~mrad.

Behind the BEMC is a time-of-flight system with a time resolution of
about 1~ns. This enables rejection of background events from
interactions of protons in the material upstream of the H1 detector.

Positrons and photons emitted at very small angles to the positron beam
direction are detected in two electromagnetic calorimeters located at
33~m and 103~m from the interaction point in the positron direction.
Designed to measure luminosity by detecting $e\gamma$ coincidences from
quasi-elastic radiative \epp collisions, they are in addition used for
studies of background arising from photoproduction.

%=======================================================================

\subsection{Event and track selection}

In this analysis two event samples are considered.  The ``low-$Q^2$''
sample consists of events in which the scattered positron is detected in
the BEMC and is limited to a range in $Q^2$ from 10 to 80~\gevsq. The
events of the ``high-$Q^2$'' sample have a positron detected in the LAr
calorimeter and have $Q^2 >$ 200~\gevsq.  The event kinematics are
reconstructed using the ``lepton-only'' method based on information from
the reconstructed positron and the event vertex.

Neutral current DIS events are selected by demanding a
well-reconstructed scattered posi\-tron with an energy larger than 12~GeV.
This ensures that the remaining photoproduction background comprises less than
1\% of the  DIS data sample.

Various other sources of background are further reduced by appropriate
selections.  To exclude events with large QED radiative effects and to
ensure substantial hadronic energy flow in the detector, the invariant
mass squared, $W^2$, of the hadronic system, determined from energy
clusters in the calorimeter, is required to be larger than 3000~\gevsq.

For comparisons with $e^+e^-$ annihilation and non-single-diffractive
data in hadron-hadron collisions, so-called ``rapidity-gap''
events~\cite{h1:rapgap} are removed from the  event sample.  To that
end, events for which the energy deposited within the polar angular
range $4.4\dgr < \theta <15\dgr$ is lower than 0.5~GeV are excluded. The
same selection is applied to the various Monte Carlo generated event
samples used in the analysis.

An event vertex, reconstructed from tracks in the Central Tracker and
located within $\pm 30$~cm of the mean vertex $z$ position, is required
to reject beam-induced background and to permit a reliable determination
of the kinematic variables.  Remaining background is rejected by
requiring no veto from the time of flight system.

%Additional selections on the high-$Q^2$ sample are applied to reject halo
%muons and to ensure $p_T$ balance. 

For this analysis, only charged particle tracks in the Central Tracker
are used, after rejection of a possible positron track. Tracks are
required to originate from  the primary vertex and to lie
within the polar angular range $15\dgr<\theta<155\dgr$.  The higher
limit eliminates the scattered positron in the low-$Q^2$ sample. In the
high-$Q^2$ data the positron track is identified with a track-cluster
linking algorithm.

With the above selection the overall efficiency for finding a genuine
primary track is better than 95\%. This estimate is based on Monte Carlo
simulations and cross-checked in a visual scan of real and simulated
events.  From this scan we further conclude that the tracker efficiency
is simulated to an accuracy of better than $2\%$.

The contamination of genuine primary tracks by tracks that have been
fitted to the primary vertex, but originate from decay products of
short lived particles, from secondary interactions and from photon
conversions, is estimated to be $10\%$, $<1\%$ and $\sim1\%$,
respectively. No spurious tracks due to noise hits have been found.

%%An additional contamination of $2\%$ is due to tracks 
%split between the inner and outer parts of CJC. 

To investigate the sensitivity of our results to the track selection
the analysis has been repeated after changing various  track quality 
criteria,
such as   the number of hits per track, the cuts on minimal radial track
length, the position in the CJC of the first measured point of a track,
the  minimum laboratory transverse
momentum and the maximum distance of closest approach to the event vertex in
the transverse plane.  The  results reported below are found to be
stable against all these variations. Residual differences are included
in the systematic uncertainties, further discussed in
Sect.~\ref{sec:syst}.

% The  rate of ``ghost  tracks'',
%spurious tracks generated by the track reconstruction algorithm, is found
%to be below 0.3\%.
%These estimates are based on Monte Carlo simulations but
%cross-checked in a visual scan of real and  simulated  events.

%To eliminate the reconstructed positron  in the low-$Q^2$ data
%from the track sample, tracks are required to have 

Within the considered kinematical regions, 53109 low-$Q^2$ and 1576
high-$Q^2$ events satisfied the above selection criteria, corresponding
to a total integrated luminosity of 1.3 \pbinv. The characteristics of
the various  event samples retained for further analysis are detailed in
Table~\ref{binstat}.

%=======================================================================

\subsection{Models for hadronic final states\label{sec:mcmodels}}

The hadronic final state of neutral current DIS events is modelled using  
several Monte Carlo event generators.  

The event sample labelled as ``DJANGO 6.0'' has been generated
with the DJANGO program~\cite{DJANGO}. It is based on
HERACLES~\cite{HERACLES} for the electroweak interaction and on the LEPTO
program~\cite{LEPTO} to model the hadronic final state. HERACLES
includes first order radiative QED corrections and the simulation of real
Bremsstrahlung photons.  The structure functions used are based on the GRV
parameterisation~\cite{GRV} and  describe the HERA $F_2$
results well.  LEPTO is used with the colour dipole model as implemented
in ARIADNE~\cite{ARIADNE} to model QCD parton cascades.  The
hadronisation phase is modelled with JETSET~\cite{JETSET} and is based on
string fragmentation~\cite{lundstring}.

The sample hereafter labelled as ``MEPS~6.4'' has also been generated with
LEPTO but  differs in the treatment of the parton cascade by using
first order matrix elements and leading-log parton showers.  It does not
include initial or final state QED radiation. For the determination of
systematic errors connected with the topology of the hadronic final
state earlier versions of the MEPS generator as well as
HERWIG~\cite{HERWIG} have been used.
  
%In all calculations, generated hadrons with 
%a lifetime larger than $8\cdot10^{-9}$~s were considered to be stable.
%These include $K^0_S$, $\Lambda$ and $\bar{\Lambda}$ particles. 

For the events generated with the models listed, the detector response was
simulated using a program based on the GEANT~\cite{GEANT}
package. The detector simulation has been checked by
comparing in detail distributions of track quality estimators with the data.

%The simulated Monte Carlo events were processed through the same
%reconstruction and analysis chain as the real events. 

The DJANGO~6.0 sample, comprising nearly 120K events after all selections,
is used here for data correction and to unfold the raw multiplicity
distributions.  The other Monte Carlo generated samples serve to
cross-check the multiplicity unfolding procedure and to determine
systematic uncertainties. 

%=======================================================================

\subsection{Data correction\label{sec:unfolding}}

The raw multiplicity distribution  in a given region of ($W,Q^2$)  and
pseudorapidity  needs to be corrected for several effects.  These
include loss of events and particles due to limited geometrical
acceptance and resolution of the tracking system, limited track finding
efficiency, contamination by tracks from particle decays and
interactions in the material of the detector which are assigned to the
primary vertex, and also QED initial state
radiation which affects the event kinematics. The results presented in
Sect.~4 are corrected for all the above-mentioned effects using the
DJANGO~6.0 event sample.
% described in Sect.~\ref{sec:mcmodels}.

Correction factors are obtained from Monte Carlo simulation by comparing
the ``true'' generated distributions before the detector simulation with
the ``observed'' distributions after this simulation followed by the same
reconstruction, selection and analysis as the real data.  The ``true''
distributions do not include the charged decay products of $K_S^0$,
$\Lambda$, $\overline{\Lambda}$ and from weakly decaying particles with
lifetime larger than $8\cdot10^{-9}\!$~s. 

%\footnote{(H1 eyes only:) Evidently, during the detector simulation,
%these particles are allowed to decay so that their decay products
%are included in the simulated track sample}.

%Most generally, both events and tracks can be lost or gained
%unwantedly.
%
%Events can be lost due to trigger inefficiencies,  event selection and  
%inaccurate reconstruction of kinematical variables.  
%The latter is important for
%events in which the positron suffers hard initial state photon radiation.
%
% Background events are non-DIS
%events but also events that migrated in the kinematical plane. 
%
%The track finding efficiency is estimated to be 
%above 95\% in the acceptance region.
%The most probable cause of track losses  is the vertex fit which  can assign
%a primary track to a secondary vertex and vice versa. Also, an acceptance
%correction will have to be applied when studying multiplicity
%distributions in centre-of-mass pseudorapidity intervals 
%lying partly outside the acceptance. 
%The acceptance region changes from event to event due to the
%boost to the hadronic centre-of-mass frame.  Finally, tracks can migrate
%in pseudorapidity and some ghost tracks may appear. 
%
%Not all of these effects can be corrected for by Monte Carlo.  However,
%the trigger is fully efficient for events with a scattered positron
%energy above 12~GeV and the purity of the final selected data sample is
%above 99\% DIS.  The remaining effects are corrected for using the 
%DJANGO~6.0 Monte Carlo event sample.
%%

The multiplicity unfolding method used here is similar to that first
employed by TASSO~\cite{tasso:mul} and subsequently adopted in various
other analyses~\cite{zeus:md,delphi:md,opal:mul,amy:mul}.  Let
$T_{true}^{MC}(m)$ be the ``true'' number of events produced by the Monte
Carlo generator with ``true'' primary charged hadron multiplicity $m$,
and $R_{true}^{MC}(m)$ the number of events with ``true'' multiplicity
$m$ observed after full detector simulation, kinematical reconstruction
and event selection.  The number of events with $m$ generated
tracks and $n$ tracks accepted after simulation and track selection,
divided by the total number of events observed with $n$ accepted tracks,
$R_{rec}^{MC}(n)$, determines a response matrix with elements $A(m,n)$.
The element $A(m,n)$ is the fraction of observed events with $n$
accepted tracks that had ``true'' multiplicity $m$.  The observed
multiplicity distribution $R_{rec}^{data}(n)$ and the ``true''
multiplicity distribution $T_{true}^{data}(m)$ are then related by

\begin{equation} 
T_{true}^{data}(m) = {T_{true}^{MC}(m) \over
R_{true}^{MC}(m)} \sum_{n} A(m,n)\, R_{rec}^{data}(n).\label{f:form}
\end{equation}

%\begin{equation} T_{true}^{data}(m) = {T_{true}^{MC}(m) \over
%R_{true}^{MC}(m)} \sum_{n}{R_{true,rec}^{MC}(m,n) \over R_{rec}^{MC}(n)}
%R_{rec}^{data}(n).
% \end{equation}

\noindent The first factor at the right-hand side of (\ref{f:form}) corrects for any
complete loss of events caused e.g. by inefficiencies and migration
outside the investigated kinematical region due to inaccurate
reconstruction of the event kinematics. 

\begin{figure}[htbp]
\begin{sideways}
\begin{sideways}
\begin{sideways}
\epsfig{figure=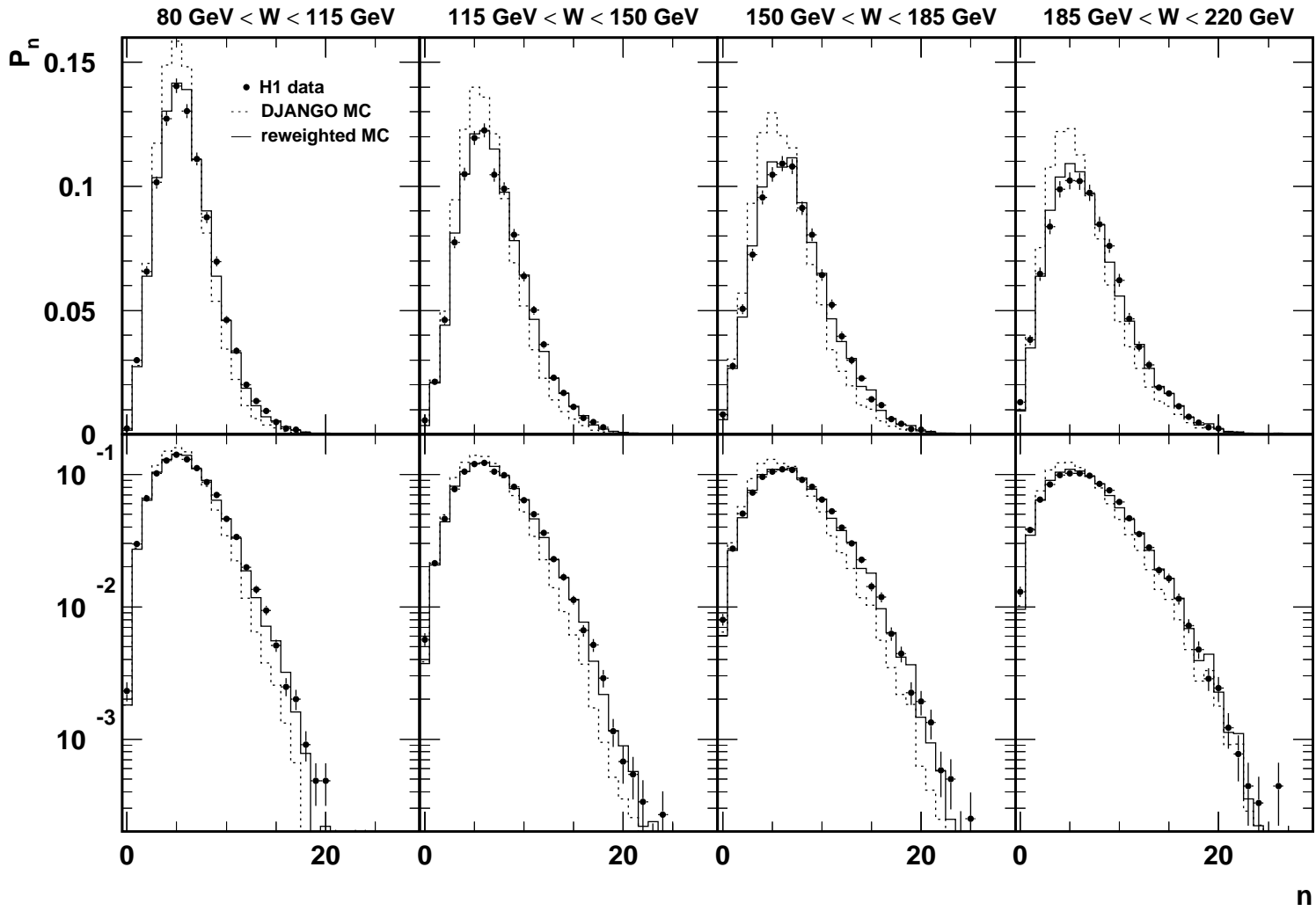}
\end{sideways}
\end{sideways}
\end{sideways}
\caption{Raw data multiplicity distributions in the 
pseudorapidity range 
$1 < \eta^\ast < 5$ compared to  
DJANGO~6.0 and reweighted DJANGO~6.0 distributions, 
after detector simulation and reconstruction, in four $W$ regions.}  
\label{fig:reweigh}
\end{figure}

One of the underlying assumptions of the method is that the relative frequencies
occurring in the response matrix derived from the Monte Carlo generator
are the same as in the real data. Otherwise, the unfolded multiplicity
distribution will be biased towards the generator input distribution. 
To reduce this bias it has been found necessary to use an iterative
procedure whereby the predicted multiplicity distribution at the
generator level is reweighted using a previous approximation of the
unfolded data multiplicity, until convergence is
reached\footnote{Convergence was reached in all kinematical regions
after 3 iterations, at the most.}.  This is illustrated in
Fig.~\ref{fig:reweigh} which shows the raw multiplicity distributions in
the interval $1 <\eta^\ast < 5$ for intervals in $W$ in a linear (upper
figures) and logarithmic scale (lower figures). The dotted histograms
are results from the DJANGO program  after detector simulation and
deviate appreciably from the real data. The result of the iterative
reweighting method is shown as the solid histogram and reproduces the
data very well.

%Two  methods are used for correction of the raw reconstructed 
%data. 
%A matrix migration method followed by a bin-by-bin correction, described
%below, has the advantage of providing a full multiplicity distribution. 
%From Monte Carlo one calculates the true number of events
%$T_{true}^{MC}(m)$ with true multiplicity $m$.  After the simulation of
%the event selection and the kinematical reconstruction one obtains the
%observed number of events $R_{true}^{MC}(m)$ with true multiplicity $m$. 
%From the simulation of track reconstruction the migration matrix
%$R_{true,rec}^{MC}(m,n)$, reflecting the observed number of events with
%true multiplicity $m$ and observed multiplicity $n$, and the observed
%number of events $R_{rec}^{MC}(n)$ with observed multiplicity $n$ are
%retrieved.  The matrix migration correction algorithm is now given by
%
%
%By dividing the matrix element $R_{true,rec}^{MC}(m,n)$ by
%$R_{rec}^{MC}(n)$, and thus calculating the probability that an event with
%observed multiplicity $n$ initially had the true multiplicity $m$, a
%dependence on $T_{true}^{MC}(m)$, the Monte Carlo input distribution, is
%introduced.

As a cross-check of the measurement of the means and dispersions of the
distributions obtained by the matrix method, and to allow the study of certain
multiplicity distributions given in parametric form, a second method is
used. Instead of correcting the observed multiplicity distribution, a
theoretical distribution $T_{true}^{th}(m,\vec{a})$, depending on the
parameter set $\vec{a}$, is transformed into a raw reconstructed
distribution taking all detector effects into account.  The function
$T_{true}^{th}(m,\vec{a})$ represents the true multiplicity distribution
for ideal non-radiative deep inelastic collisions.  The transformed
distribution $R_{rec}^{th}(n)$ is then compared to the measured one and the
parameters $\vec{a}$ determined with a standard minimum $\chi^2$ procedure. 

A more detailed discussion of the merits and limitations of this method
is given in~\cite{e665:mul,delphi:mul}. 
In the present paper it is used to make comparisons with
the Negative Binomial and Lognormal distributions.
 
%=======================================================================
 
\subsection{Systematic errors\label{sec:syst}}

In this section we discuss various sources of possible systematic
uncertainties. Each possible systematic effect is independently varied
and the analysis repeated as outlined in the previous section.  The
difference in the final result, relative to the quoted result based on 
DJANGO~6.0, is taken as one contribution to the systematic uncertainty.
The errors from all sources are added in quadrature.

%The largest systematic uncertainty in our results originates from the
%acceptance limitation of the Central Tracker and the corresponding
%corrections using Monte Carlo models.  This is a consequence of
%relatively large differences in the predictions for hadronic final
%states obtained by present-day models for DIS.  This source of
%systematic error especially affects the measurements at low $W$ and in
%the full current hemisphere ($\eta^\ast>0$) discussed in
%Sect.~\ref{sec:curr}. It is, however, much less important for the
%restricted pseudorapidity domains studied in Sect.~\ref{sec:subd}.

For illustration, the change  in the mean charged
multiplicity due to various effects are summarised in
Table~\ref{syseffects}. Results are given for both the full current
hemisphere ($\eta^\ast>0$) where the corrections are largest
 and the pseudorapidity range $1 < \eta^\ast <
3$, where the detector acceptance is high.

\begin{itemize}

\item Photoproduction events that survive the DIS selection will have a
fake positron detected in the BEMC.  Above a reconstructed scattered
lepton energy $E'_e$ of 14~GeV the contribution of photoproduction
events is practically zero.  For $12 < E'_e < 14$~GeV the background
contribution is estimated to be  5\%~\cite{h1:f2_96}.
This means that photoproduction events  only contribute in the
highest $W$ interval. To exclude  events in this interval 
the $E'_e$ cut is raised
from 12 to 14~GeV, a smaller radial extension of the BEMC energy cluster
is demanded and events with a positron detected in the electron tagger
are rejected. This results, however, in a negligible change of the mean
multiplicity in the highest $W$ interval.

\item The BEMC energy scale is known to an accuracy of $1\%$.  
Decreasing   the energy scale by its error  reduces
 the mean multiplicity by a maximum of $\sim 3\%$. 

%\item The DJANGO Monte Carlo, which is used for correction, simulates
%radiative effects.  To lower the influence of radiative events the cut on
%$W^2$ determined from energy clusters was raised from 3000~\gevsq\ to
%4400~\gevsq. 

%
%\item Kinematical variables are reconstructed using the lepton-only
%method.  An alternative reconstruction method, using hadron information
%only, gives deviations of around 2\%. 

\item The track selection criteria ensure a high efficiency (above 95\%
inside the acceptance region). To restrict tracks to a region of even
higher efficiency additional  kinematical selections are introduced. The
polar angle $\theta$ of a track  is required to be larger than 22$\dgr$
and the transverse momentum $p_T$ to be larger than 0.15~GeV/c. After
correction, this decreases the mean multiplicity in the full current hemisphere
by about 2\%.  However, exclusion of tracks from the forward Central
Tracker region or with low $p_T$ enhances the sensitivity to the type of
Monte Carlo generator used. The contribution to the total systematic
error is, therefore, estimated by applying additional selections on the
track {\em quality}, requiring more than 10 hits, a track length of more
than 10~cm and a start point in  CJC1. With these selections, the  fraction
of  tracks for which track segments in the inner and outer
CJC are not linked, and therefore counted double, is kept 
below $0.3\%$ and reproduced by the Monte Carlo simulation within $30\%$.

%The ghost rate which is
%completely due to split tracks is described by the Monte Carlo simulation
%within 0.1\%. 

%Recently it was discovered that the central jet chambers were slightly
%twisted.  As a result of this the CJC has to be recalibrated and the whole
%data sample will be reprocessed.  It was possible however, to provide a
%crude correction so that the effect of this recalibration could be
%estimated.  Since a multiplicity study doesn't require an extremely
%accurate momentum reconstruction, but rather relies on the track finding
%efficiency, the influence of the recalibration is minor.  A change in mean
%multiplicity of less 0.2\% is expected, which is well below the
%statistical error. 

\item The multiplicity unfolding is performed iteratively to reduce the
dependence on the generated Monte Carlo multiplicity distribution. To
estimate the importance of any residual systematic bias on the mean and
dispersion of the multiplicity distribution, a Negative Binomial
distribution, smeared for detector effects, is fitted to the
reconstructed data multiplicity distribution. The effect on the
determination of the mean multiplicity is well below $2\%$. 
In addition, $\avgn$ has been
estimated from the fully corrected single-inclusive charged particle
$\eta^\ast$ spectrum in each $W$ interval, using the standard bin-by-bin
correction method. The (generally small) difference between extremes is
taken as one contribution to the overall systematic error.

\item 
The predictions for $\aver{n}$ of the Monte Carlo generators  used here differ
by $17\%$, at most, in the  pseudorapidity domain $0<\eta^\ast<1$.
This region lies partly outside the Central Tracker acceptance at low $W$.
Since a change in generator multiplicity outside the acceptance region does
not affect the reconstructed distribution, a generator dependent bias 
cannot be removed by reweighting the input distribution. 
A systematic uncertainty has been assigned to the results pertaining to the
full current hemisphere ($\etas>0$) by taking
the maximum difference  between results derived from different event
generators. Table~2 illustrates the maximum size of the effect
obtained from the generator LEPTO~6.1. 
%Once again, the point raised in this
%paragraph does not apply to the %multiplicity in restricted pseudorapidity
%intervals. 

\end{itemize}

%Diffractive events are removed from the DIS sample by rejecting events
%with a ``rapidity gap''.  {\it Not} removing these events would result in
%a 3\% decrease of the mean multiplicity. This was not included in the
%quoted systematical errors. 

The contribution from decay products of $K^0_S$, $\Lambda$,
$\bar{\Lambda}$ etc.  is subtracted via the unfolding procedure. 
However, recent studies at HERA indicate that the $K^0_S$ production
rate may be overestimated by about 10\%~\cite{zeus:kzero,h1:kzero} in the  models
used for correction.  Such a difference leads to a systematic
underestimation of the mean multiplicity of at most 0.4\%.

From event simulation it is estimated that about 0.1~tracks per event
assigned to the primary vertex  are, in fact, due to photon conversions
in the detector material. To account for possible differences between
the event simulation and the real detector response an additional 1\%
systematic error on the mean multiplicity is assumed.

%As a further check, average multiplicities were also obtained from the
%inclusive $\eta^\ast$ distribution, corrected via a standard, Monte Carlo
%based bin-by-bin correction procedure. We find very good agreement in all
%pseudorapidity domains for $\eta^\ast > 1$. 

%=======================================================================

\section{Results}

The  multiplicity distribution has been measured, in the kinematic
regions in $W$ and $Q^2$ listed in Table~1, for charged particles with
pseudorapidity in the domains $1\leq\eta^\ast\leq\eta^\ast_c$ with
$\eta^\ast_c=2,3,4,5$ and in intervals of unit pseudorapidity centered at
$\eta^\ast=2.5,3.5,4.5$, as well as for the full current hemisphere
defined as the domain $\eta^\ast>0$. The data are  integrated over
particle centre-of-mass transverse momentum to allow comparison with
other experiments.

The corrected multiplicity distributions, in four $W$ intervals
and four $\eta^\ast$ domains, are given in Table~3. For each
multiplicity, $n$, the relative frequency, $P_n$, the statistical
uncertainty (first) and the systematic uncertainty (second) are given.
The statistical uncertainties take into account the finite number of
events in the data and in the generated samples. They are calculated by means
of a Monte Carlo sampling procedure which ensures propagation of sampling
fluctuations and statistical correlations into the statistical error of
the measurement. The same technique is employed to determine statistical
errors on   moments of the \MD{} and other derived quantities. It must
be noted that the $P_n$ values given in the table are correlated for
nearby values of $n$ due to the method of correction.

%In these data, the charged decay products of $K^0_S$, $\Lambda$ and
%$\overline{\Lambda}$ are excluded.
 
Table~4 summarises corrected results in various pseudorapidity domains
for the mean charged
multiplicity $\avgn$, the generalised dispersions $D_q$, the normalised
multiplicity moments 
%\footnote{Note (see e.g.~ref.~\cite{ua5:md:2:9} for
%explicit formulae) that all multiplicity moments, 
%and their errors, for
%a given distribution are highly correlated} 
$C_q$ ($q=2,3,4$) and the
factorial moments $R_2$, $R_3$. The table also lists values for
the  third order factorial cumulant $K_3$  
for those intervals where a statistically significant measurement
is possible.

Estimates of the first two moments (i.e. $\avgn$ and $D$) of each
multiplicity distribution have also been obtained using the parametric
method, described in Sect.~3.4, based on fits of the Negative Binomial
and Lognormal distributions. The results from this  method  and from 
the  matrix unfolding are found to be consistent.

Not all of the statistical moments  listed in Table~4 will be discussed
in detail. They have been measured in many experiments and are included
for reference  purposes.

%~\footnote{All H1 data presented in this paper are
%also available   from the authors}.

%=======================================================================

\subsection{Current hemisphere multiplicity moments\label{sec:curr}}

To allow comparison with  lepton-nucleon  fixed target
data and single hemisphere data from $e^+e^-$ experiments, we present in this
section results on the moments of the charged particle multiplicity
distribution in the full current hemisphere  $\etas>0$. The limited
experimental acceptance in the interval \mbox{$0<\etas<1$}, in particular
for  the range  $80<W<115$~GeV, renders the H1 data in this region more 
sensitive to the Monte Carlo generators used in the correction procedure
than in other  pseudorapidity domains. The quoted systematic errors
reflect  this  additional uncertainty\footnote{ Diffractive events are
removed from the H1 DIS sample by rejecting events with a ``rapidity
gap''.  {\it Not} removing these events would result in a 3\% decrease
of the mean multiplicity. This effect is not included in the quoted
systematical errors.}.
  
Data on multiplicity distributions in a single hemisphere and in
restricted domains of rapidity   are available from several $e^+e^-$
experiments~\cite{delphi:md,opal:mul,tasso:mul,hrs:mul}. 
To compare  DIS data with that for $e^+e^-$
annihilations we have chosen to use the $e^+e^-$ JETSET parton shower model,
with parameter settings as used by the DELPHI
collaboration~\cite{delphi:jet:tune}. This model reproduces 
in  detail the \MDS\ in the full phase space and in
restricted intervals of rapidity~\cite{delphi:md,aleph:rap:mul}.

%including the so-called intermittency effect~\cite{review:93b}.
%HERWIG, with cluster fragmentation, also describes the
%MD data, but less well~\cite{opal:mul}.

The JETSET $e^+e^-$
predictions presented below (labelled as ``JETSET $e^+e^-$'' in the figures)
are obtained for a mixture of ``primary'' light quark pairs
only. Decays of $K_S^0$, $\Lambda$ and $\overline{\Lambda}$ are treated
as in the DIS data and in the simulations, thereby avoiding experiment
and energy dependent corrections to published data\footnote{We have
verified that the JETSET $e^+e^-$ predictions, with five active
flavours, reproduce the published data in the PETRA-LEP energy range,
including recent measurements at 130 GeV~\cite{delphi:md:130}.}. 
The contribution to $\aver{n}$ from charmed quark fragmentation,
estimated at LEP to be a factor $1.02\pm0.03$ larger than for light
quarks~\cite{delphi:charm} is neglected.
The model predictions are also  used  in domains of  pseudorapidity  where
no direct  $e^+e^-$ measurements exist. For these, the conclusions
should be treated with caution.

%=======================================================================

\subsubsection{Mean Charged Multiplicity}

Figure~2a  shows the mean charged multiplicity for $\etas>0$, measured
by this and other lepton-nucleon
experiments~\cite{jones:92,e665:mul,jongej:89,emc:mul} as a function of
$W$.

%Also included is a data point for the meson
%fragmentation region in the centre-of-mass 
%of non-single-diffractive $\pi^+p$ collisions 
%at $W\equiv\sqrt{s}=21.7$ GeV~\cite{na22:neg:binom1},
%the highest energy reached so far in meson-hadron interactions.

Here and in the following, unless stated otherwise, the total errors are
the overall uncertainties  computed by adding the statistical and
systematic errors in quadrature. When two error bars are displayed the
inner error bar is the statistical error and the outer one is the total
error. For data from other experiments we have used the published
systematic errors, whenever available. Otherwise a  systematic error of
$5\%$ is assumed.

In the $W$ range  covered by HERA, $\avgn$ is compatible with a linear
increase with $\ln{W}$. Combined with the   data  at   lower
energy\footnote{For clarity, only  a representative sample of $\nu p$
data is plotted.} it is evident, however,  that the mean multiplicity
increases faster than $\ln{W}$. The HERA data confirm, for the first
time in DIS  lepton-proton scattering, the faster-than-linear
growth of $\avgn$ with $\ln W$, a feature already well-known from $e^+e^-$
annihilations and hadron-hadron collisions, and expected in  perturbative
QCD.

Various models predict  the evolution  of the mean multiplicity with
energy. We have fitted\footnote{Because of the systematic difference
between the EMC and E665 results, discussed in~\cite{e665:mul}, fits
are performed {\em without} the E665 data points. The errors on
best-fit parameters quoted in this section include systematic
uncertainties.} several parameterisations to the data plotted in Fig.~2a.

%The power-law form
%\begin{equation}
%\AVN= a\cdot (W/W_0)^{2b},  \label{f:fermi}
%\end{equation}
%originally obtained by Fermi~\cite{heisenberg:fermi} was also derived in
%hydrodynamical and
%fireball models~\cite{satz}, 
%and in scale-invariant branching processes with energy independent 
%coupling~\cite{scale:cascade}.
%The fit, with $W_0=1$ GeV,  yields 
%$a=1.05\pm0.04$, $b=0.226\pm0.005$ with a $\chi^2$ per degree
%of freedom $\CHDF= 2.3$.

According to the KNO-G prescription an appropriate power-law form is,
following~\cite{wrob:logn}
\begin{equation}
\avgn = a \cdot (W/W_0)^{2b'} -c, \label{f:wrobn}
\end{equation}
\noindent where the constant $c$ may be regarded as a ``discreteness
correction'' with the value $c=0.5$ and $W_0=1$ GeV. The fit yields $a=1.40\pm0.04$,
$b'=0.20\pm0.01$ with $\CHDF=39/23$. 
The value for  $b'$ agrees with that
obtained by Szwed {\em et al.\/} ($b'=0.221$)
%result with the values $a=2.69$, $b=0.221$ obtained by Szwed {\em et
in  a comprehensive analysis of the {\em full} phase space
multiplicity distribution in $e^+e^-$ annihilation~\cite{wrob:logn}.

%A fit to a form motivated by perturbative QCD calculations of the
%parton evolution in the leading-log approximation~\cite{furmanski}:
%\begin{equation}
%\AVN= a\ +b\,\exp{(c\sqrt{\ln(W^2/Q_1^2)}}  , \label{f:furmanski}
%\end{equation}
%with  $Q_1=1$ GeV
%yields $a=1.47\pm0.06$, $b=0.046\pm0.003$,
%$c=1.66\pm0.02$ with $CHDF=3.7$.
%and $c$ fixed at its theoretical value 
%$c=\sqrt{72/(33-2N_f)}=1.697$  for $N_f$, the number of active flavours, 
%$c=\sqrt{72/(33-2N_f)}=1.633$  for $N_f$, the number of active flavours, 
%equal to three,
%yields $a=1.25\pm0.05$, $b=0.040\pm0.003$  with $\CHDF=3.7$.
%These results may be compared  to  the values 
%  $a=1.43\pm0.12$, $b=0.046\pm0.003$
%for $\mu^+ p$ current hemisphere data in the range $5<W<20$ GeV~\cite{emc:85np}.
%This parameterisation, however, is  statistically incompatible with the DIS
%experimental results  and grows faster than the data in the HERA region.

%In $e^+e^-$ annihilation, the form (\ref{f:furmanski})
%was used in several studies~\cite{opal:mul,tasso:mul,pluto:mul,amy:mul}.
%As a representative example, we quote the   OPAL result~\cite{opal:mul}
%for $\AVN/2$: $a=1.209\pm0.042$, $b=0.056\pm0.006$, $c=1.712\pm0.035$
%($\CHDF=94/81$), comparable with the DIS result.

We have further compared the DIS data to the Modified Leading-Log (MMLA+LPHD)
prediction~\cite{doksh:91} in the form proposed in~\cite{lupia:ochs} and
valid for running QCD coupling $\alpha_s$:

%\footnote{%
%See ref.~\cite{doksh:91} eqn.~7.32}

\begin{equation}
\avgn = c_1 \, \frac{4}{9}\,N_{LA} + c_2, \label{f:lupia:ochs}
\end{equation}
\noindent with
\begin{equation}
N_{LA}(Y)=\Gamma(B)\,\left(\frac{z}{2}\right)^{1-B}\,I_{1+B}(z),\label{f:mlla}
\end{equation}
 where $z \equiv \sqrt{\frac{48}{\beta_0}Y}$, with $Y=\ln(W/2Q_0)$,
$a=11+2N_f/27$, $\beta_0=11-(2/3)N_f$, $B=a/\beta_0$ and $N_f$
the number of active flavours; $I_\nu$ is a
modified Bessel function of order $\nu$ and $\Gamma$ is the Gamma
function.

In~\cite{lupia:ochs} this expression was shown to describe the mean
charged multiplicity in $e^+e^-$ annihilation from LEP energies down to
centre-of-mass energies of $3$ GeV. The factor $\frac{4}{9}$ accounts for the
multiplicity difference in  a quark and gluon jet; $c_1$ is a
(non-perturbative) normalisation parameter and $c_2$ the ``leading
parton'' contribution, not included in the theoretical calculation.
Using the same shower cut-off value $Q_0=270$ MeV and  $N_f=3$ as
in~\cite{lupia:ochs}, we find $c_1=1.21\pm0.05$, $c_2=0.81\pm0.08$ with
$\CHDF=45/23$. The best-fit curve is shown in Fig.~2a  (dashed line) and
describes the data over a wide $W$ range.

%An expression nearly identical to (\ref{f:mlla}) 
%has  also  been derived within the Lund dipole cascade formalism~\cite{Dahl89}.
%It was shown to follow closely results on hadron multiplicities
%obtained  from the ARIADNE generator.

The mean multiplicity has also been computed as a function of
$\alpha_s(W)$ including the resummation of leading and next-to-leading
corrections~\cite{webber:84} with the result
\begin{equation}
\avgn = a\, \alpha_s^b \exp{(c/\sqrt{\alpha_s)}}
\, \left[1+d\cdot\sqrt{\alpha_s}\right],   \label{f:webber}
\end{equation}
\noindent where the parameter $a$ cannot be calculated from QCD.
The constants $b$ and $c$ are predicted by theory\footnote{%
Note that formula (\ref{f:mlla}) reduces to (\ref{f:webber}) at large
$z$.}. This QCD prediction has been successfully tested over a wide
energy range in several analyses of the mean charged multiplicity in $e^+e^-$
annihilation~\cite{opal:mul,aleph:mul,delphi:md,amy:mul}, including the
recent LEP measurement at 130~GeV~\cite{delphi:md:130}. For the running
coupling constant we use the two-loop expression
\begin{equation}
\frac{\alpha_s(W^2)}{4\pi}=\frac{1}{\beta_0\ln(W^2/\Lambda^2)} -
\frac{\beta_1\ln\ln(W^2/\Lambda^2)}{\beta_0^3\ln^2(W^2/\Lambda^2)},
\label{f:twoloop}
\end{equation}
\noindent with the constant $\beta_0$ as defined before,
$\beta_1=102-38N_f/3$, $b=1/4 +(10N_f)/(27\beta_0)$ and
$c=\sqrt{96\pi}/\beta_0$. According to~\cite{webber:84} $\Lambda$ needs
not be identical to $\Lambda_{\overline{\rm MS}}$, although both are
expected to be rather similar, in particular if the O$(\sqrt{\alpha_s})$
correction turns out to be small.

A fit of the data to the form~(\ref{f:webber}) with $a$ and $d$ as free
parameters, $\Lambda=263$ MeV~\cite{virchaux} and $N_f=3$ yields
$a=0.041\pm0.006$ and $d=0.2\pm0.3$ with $\CHDF=27/23$.
Neglecting the $O( \sqrt{\alpha_s} )$ correction in (\ref{f:webber}) and 
treating $\Lambda$ as a free parameter, we find
$\Lambda=190\pm60$~MeV and $a=0.034\pm0.005$ with $\CHDF=29.9/23$.
The functions (\ref{f:wrobn}), (\ref{f:lupia:ochs}) and
(\ref{f:webber}) are fairly similar\footnote{The corresponding fitted
curves practically coincide with the dashed line in Fig.~2a and are not
shown for reasons of clarity.} and clear differences between them will become
visible only at much  larger energies.

%In the latter experiment,   a fit to all data
%above the $b\overline{b}$-threshold and using $N_f=5$,
%yields  $a=0.057\pm0.006$, $d=0.43\pm0.35$.
%The fit to {\em single hemisphere}  DIS data, using $N_f=4$
%compares favourably with this global $e^+e^-$ fit\footnote{%
%It should be remarked  that the  sensitivity 
%of eqn.(\ref{f:webber}) to $N_f$ is very weak
%since in QCD the multiplicity is determined primarily by the dynamics of
%gluons, not quarks. The number of flavours enters mainly via
%relatively unimportant loop corrections}.

The above comparisons confirm and extend earlier indications from $\mu
p$ interactions~\cite{emc:85np} that the rate of increase with energy of
$\avgn$ in the current fragmentation region of DIS lepton-nucleon
interactions  is similar to that observed in $e^+e^-$ annihilation in the
presently covered energy range.

The energy evolution of the mean multiplicity of partons emitted from a
primary parton has been calculated 
%in the Double Logarithmic Approximation (DLA) 
for running as well as for fixed
$\alpha_s$~\cite{doksh:91,ochs:vietri}. In the latter case the
multiplicity rises as a power of the energy as in
equation~(\ref{f:wrobn}). For running coupling the
growth is slower than any power of $W$, but faster than any power of
$\ln{W}$, as in equations (\ref{f:lupia:ochs}) and
(\ref{f:webber}).

The various parameterisations discussed here show that, 
up to present energies, distinction between fixed and
running  $\alpha_s$, based on measurements of $\avgn$, is still not
possible. The data are sensitive, however, to soft gluon interference in
QCD. Neglecting interference would increase the multiplicity anomalous
dimension~\cite{doksh:91}, $\gamma={\rm d} \ln \avgn / {\rm d} \ln W^2$, by a factor
$\sqrt{2}$. This has been shown~\cite{ochs:vietri} to be inconsistent
with data for any reasonable value of $\Lambda$ and
confirmed for DIS in~\cite{zeus:md,h1:thompson}.

In a strict sense the perturbative QCD results apply only to
the soft component of the parton cascade emitted by a single quark or
gluon jet. They are approximately valid for initial parton
configurations such as those encountered in $e^+e^-$ annihilation and in the
simple quark-parton picture of DIS. However, the dynamics of DIS
processes is more complex than  in $e^+e^-$ and differences of detail
are to be expected~\cite{kant:thesis}.

%Nevertheless, in the context of the presumed universality of quark
%
In Fig.~2a we  compare the DIS multiplicity
with single hemisphere results as expected from the JETSET generator
 in 
$e^+e^-$ annihilation for light quark-antiquark 
pairs (dotted line)\footnote{The single hemisphere
multiplicity distribution for $e^+e^-$ is calculated relative to the thrust
axis.}.  
The  evolution with energy of $\aver{n}$ in $e^+e^-$
is similar to that in DIS. 
The H1 results are consistent with the presence of a  small 
multiplicity excess in $e^+e^-$ annihilation, relative to DIS, above
$W=10$ GeV, which has  been noted before~\cite{emc:ee,e665:spectra:91}.
In measurements of inclusive charged particle spectra at
HERA~\cite{h1:eflow:94,zeus:single:95} a similar excess is seen near
zero longitudinal momentum  in the hadronic centre-of-mass. It is
usually attributed to more prolific gluon emission in $e^+e^-$. The
MEPS~6.4 generator for DIS at the HERA energy (solid line) overestimates
substantially the mean charged multiplicity.

%=======================================================================

\subsubsection{The ratio $\avgn/D$ and $R_2$}

In Fig.~2b we show the ratio of $\avgn$ to $D$  in a comparison with
fixed target DIS data. This ratio is expected to be energy
independent if KNO scaling holds. The data above 10 GeV are indeed
constant, within large errors, after a  clear rise at lower $W$. The
JETSET $e^+e^-$ prediction (dotted curve) exhibits  rather  similar energy
dependence and, moreover, illustrates that KNO scaling is  only
approximately valid. The MEPS~6.4 generator  (solid line) agrees
well with the HERA data.

The normalised second order factorial moment $R_2=\aver{n(n-1)}/
\avgn^2$ is plotted in Fig.~2c. This quantity is equal to the integrated
two-particle inclusive density and is, therefore, a direct measure of
the strength of hadron-hadron correlations. It shows little, if any, 
energy dependence over the HERA range but rises steadily at lower $W$.
The behaviour is very similar for $e^+e^-$ final states. The 
experimental values of $R_2$ are further compared with a QCD calculation
which, for a quark jet, predicts $R_2$ to behave as

\begin{equation}
R_2=\frac{7}{4} \,\left[1-\kappa\sqrt{\alpha_s}\right],\label{f:f2qcd}
\end{equation}

\noindent with $\kappa=0.88$ for three flavours~\cite{malaza:webber}.
Leading and next-to-leading order  predictions are plotted, with
$\alpha_s$ calculated according to the two-loop formula
(\ref{f:twoloop}) and the scale parameter set to $\Lambda=263$ MeV. Both
curves are significantly above the data. Nevertheless, it is interesting
that  the  next-to-leading order calculation  comes closer to the data,
although the disagreement  remains considerable. 
The data are rather well reproduced by the  JETSET
model in the case of
$e^+e^-$ annihilation, a fact confirmed by differential 
measurements 
of the two-particle correlation function by OPAL~\cite{opal:corr:xi}. The
prediction of MEPS~6.4 for DIS (solid line) is in agreement with the HERA
results.

The results presented in this and the previous section indicate similarities
of the low order moments of the single hemisphere multiplicity distribution
in DIS and $e^+e^-$ annihilation,
in conformity with the hypothesis of approximate universality
of quark and gluon fragmentation.

QCD predictions for the energy dependence of the mean {\em parton}
multiplicity, derived from analytical solutions of the evolution
equations, are in agreement with that observed for {\em hadrons} and
thus add to the existing  support for the LPHD {\em ansatz} at the
single-inclusive level. The  large disagreement  between data and QCD
next-to-leading order calculations of~\cite{malaza:webber} 
for $R_2$  suggests that extension of the LPHD hypothesis to  higher order
inclusive correlations may not be justified~\cite{ochs:vietri}.

%The  observed similarities of DIS and $e^+e^-$ multiplicity moments are
%somewhat surprising  since well-known DIS processes such as
%initial state colour radiation, remnant fragmentation and boson-gluon
%fusion, particularly important at small Bjorken-$x$, have no analogues
%in $e^+e^-$. To reveal such differences a more differential analysis of
%the \MD{} is required.

%For $e^+e^-$, both the quark and the antiquark
% can radiate gluons whereas in DIS, radiation from the
% target remnant---a composite colour source---is strongly suppressed.
% The probability per event for gluon radiation is consequently
% larger in $e^+e^-$.
%
% A priori one must expect  differences of detail in the
% structure of hadronic final states for the two types of collisions.
% 
% For a fully quantitative comparison detailed models are needed which
% contain all the theoretical ingredients outlined above.
% {\bf[comment on   theoretical e.g. BGF uncertainties]}
% %
% It is instructive, nevertheless, to compare 
% experimental data for the two processes as directly as feasible. 
% Models simulating the hadronic final states of
%  soft hadron-hadron interactions, such as the Dual-Parton
% model~\cite{dpm}
% FRITIOF~\cite{fritiof} or PYTHIA~\cite{pythia}, have not yet
% reached the level of sophistication of those simulating
% hard processes and are no alternatives for direct data comparisons.

%=======================================================================

\subsection{The multiplicity distribution in pseudorapidity
domains\label{sec:subd}}

In this section we study the \MD{} and its moments in domains of
pseudorapidity, limited to the current fragmentation region $\eta^\ast>1$. 
We examine the $Q^2$ and $W$ dependence and compare 
with data from other types of interactions and with predictions
from the MEPS~6.4 generator.

%=======================================================================

\subsubsection{$Q^2$ dependence}

In the simple quark-parton model the properties of the total hadronic
system produced in a deep inelastic lepton-hadron collision depend on
the lepton kinematical variables Bjorken-$x$ and $Q^2$ only through the
invariant mass $W$ of the hadronic  system. In QCD, scaling violation of
the quark fragmentation functions and of the parton distributions
introduce an explicit $Q^2$ dependence even at fixed $W$. Fixed target
DIS electron, muon and (anti)neutrino experiments at low energies
confirm that the global characteristics of the hadronic final states and
the average number of produced  hadrons in particular, vary most
significantly with $W$. At fixed $W$,  only weak
dependences on $Q^2$ are observed~\cite{schmitz:rev,noqvar}. 
Such  results are in
accord with the Bjorken-Kogut correspondence principle~\cite{bj:kogut}
and imply that the densities in the hadron plateaus spanning the
current,  central and target regions are quite
similar~\cite{bj:kogut,cahn:cleymans}. Only recently has a statistically
significant $Q^2$ dependence of the mean charged hadron multiplicity
been established in $\mu^+p$ and $\nu(\overline{\nu})p$
interactions~\cite{emc:86,jones:91}. The effect is limited to a
restricted region in Feynman-$x$: $-0.15<x_F<0.15$, where $x_F$ is the
fractional longitudinal momentum of a hadron in the hadronic
centre-of-mass frame. There are no published results on a possible
variation with $Q^2$ of  the {\em shape} of the multiplicity
distribution.

The H1 experiment  allows us to investigate the charged particle
multiplicity distribution over a widely extended range of $Q^2$ in a
novel energy domain. In Fig.~3 are plotted the mean charged multiplicity
and the dispersion of the \MD\ in four intervals of $W$, and in the
pseudorapidity domain $1<\eta^\ast<5$, covering part of the current
fragmentation hemisphere. Within errors, no significant variation with
$Q^2$, in a fixed interval of $W$, is observed for $\AVN$ and $D$ in the
range $10<Q^2<1000$ GeV${^2}$. The solid lines in Fig.~3  are fits to a
constant.

To ascertain the evolution with $Q^2$ seen at much lower energy it will
be necessary to study the multiplicity distribution in the central and
proton remnant pseudorapidity regions which are not covered in this
analysis. A similar study with quasi-real photons at HERA would be of
evident interest.

The data presented in subsequent sections have been obtained from data
samples averaged over the full $Q^2$ region ($Q^2>10$ GeV${}^2$) covered
by this experiment.

%=======================================================================

\subsubsection{The shape of the multiplicity distribution}

In Fig.~4 we show, as a representative  example, the multiplicity
distribution in the interval $115<W<150$ GeV, measured  in various
$\eta^\ast$ domains. In this  figure only statistical errors  are
plotted. The distribution for the widest $\etas$ interval is plotted at its
true scale. The distributions for the other intervals are successively shifted 
down by a factor ten.

The figures illustrate that the multiplicity distribution, at fixed $W$,
becomes narrower as the size of the $\etas$ interval is reduced.
However, the same distributions plotted in KNO form (not shown)
widen under the same conditions. The latter property is
in part related to the diminishing influence of global conservation
constraints and was first predicted in~\cite{bialas:hayot}.

Figure~4 also shows the predictions from MEPS~6.4  (open symbols).
Significant deviations are noted. The  model overestimates the mean
multiplicity (cf.~Fig.~2a) and, consequently, does not reproduce the
small and high $n$ tail of the distribution. These defects  are seen in
all $W$ and $\etas$ intervals examined.

Numerous parameterisations for the shape of the multiplicity
distribution are used  in the literature. Here we concentrate on the
Negative Binomial Distribution (NDB) and the Lognormal Distribution
(LND). The phenomenological arguments leading to these forms are
discussed in Sect.~\ref{sec:feno:qcd}. The parameters of these
parametric models are obtained from a least $\chi^2$ fit to the
uncorrected multiplicity distributions, as explained  in Sect.~3.4.
%
%The  data point at zero multiplicity ($n=0$) is not
%measured in the interval $1<\eta^\ast<5$, was not included in the fits.
%
The best-fit parameters for all studied pseudorapidity and $W$ intervals
are summarised in Table~\ref{tab:fitpar}. The  errors quoted  are the
quadratic sum of the statistical error and the systematic uncertainties.

The solid (dashed) line in Fig.~4 shows how the LND (NBD) compares to
the measurements. Inspection of this figure and of
Table~\ref{tab:fitpar} indicates that the LND  gives a reasonably
accurate description of the data, in particular  in the smallest
$\eta^\ast$ domain. However, the quality of the  fits deteriorates in
larger domains. Likewise, the NBD fits are acceptable 
in the smallest $\eta^\ast$
domain but become progressively worse for larger intervals. The two
distributions are seen to differ most for low multiplicities.
Nevertheless, it may be verified from Tables 4 and 5 that the estimates
for the mean and dispersion of the multiplicity distribution, derived
from these parameterisations, agree very well with those derived from
the fully unfolded distribution. Parametric forms, such as the NBD and
LND, therefore  remain useful for phenomenology.

In intermediate size rapidity intervals in $e^+e^-$ annihilations at LEP both
the LND and the NBD are unable to describe the multiplicity distribution
which exhibits a prominent shoulder at intermediate $n$ values. The
shoulder is most prominent in the single hemisphere
distributions~\cite{delphi:md,aleph:rap:mul}.
This structure results from  a superposition of two-jet
and three- or four-jet events. The effect demonstrates that the
fluctuations in  the number  of hadrons, and therefore the \MD{},
carries information on the hard partonic  phase of the multihadron
production process, even after soft hadronisation. The  connection between
parton level and hadron level dynamics has
been investigated  by the Lund group for parton cascades treated in the
dipole formalism~\cite{Dahl89,lund:anom}. It is shown that, for
centre-of-mass energies above $~\sim50$ GeV, the fluctuations in the
{\em hadron} multiplicity are to better than $90\%$ determined by the
hardest and  second hardest gluons emitted. Further softer radiation and
subsequent (string) hadronisation, adds only small
(sub-Poissonian) fluctuations to those induced in the initial stage of
shower development. These analytical results provide a quantitative
realisation of the notion of Local Parton-Hadron Duality derived
directly from  perturbative QCD.

The H1 measurements show no evidence for a shoulder structure of
the type seen in $e^+e^-$. Neither is such a structure present in the
multiplicity distribution predicted  by MEPS~6.4.
An excess of high-multiplicity events would also be expected
if a significant proportion of DIS events were induced by 
QCD instantons~\cite{ringwald}.
In order to determine from our data an upper limit for the cross
section of such events, we closely follow 
an analysis method recently applied by H1
in a study of strange particle production~\cite{h1:kzero}.
We assume that the observed multiplicity distribution
is a superposition of two distributions, one associated with
instanton-induced events, another with ``standard'' DIS events.
The former is calculated from the instanton generator described
in~\cite{h1:kzero}. For the latter we adopt a Negative Binomial
form.
Using a $\chi^2$ minimisation procedure to determine the
relative proportion of instanton-induced and standard DIS events,
we derive a $95\%$ confidence level upper limit of $0.3$~nb on the
cross section for instanton production in the pseudorapidity domain
$1<\eta^\ast<5$  and $80<W<115$~GeV.

The H1 data, which cover larger $W$ values than presently reachable in
$e^+e^-$ annihilation, are  in qualitative accord with the
QCD expectation that hard multi-jet production is less frequent
in lepton-hadron collisions. Radiation from the target remnant---a
composite colour source---is expected to be strongly suppressed by 
the so-called ``antenna effect''~\cite{lund:anom}. For sufficiently small
values of Bjorken-$x$ it  also  suppresses hard gluon radiation
in the current region. The probability per event for hard
gluon radiation is consequently larger in $e^+e^-$ annihilation.
As shown in~\cite{lund:anom}, the antenna effect  leads to  
smaller values of the {\em local} multiplicity anomalous dimensions  and to
smaller  multiplicity fluctuations for DIS  in the current region, 
compared to $e^+e^-$ annihilation.
These topics  are  examined  in the following sections.

%=======================================================================

\subsubsection{KNO scaling and correlations\label{sec:kno}}

To demonstrate  the energy scaling of the multiplicity distribution in
DIS in the  energy range opened up by HERA we show in Fig.~5 the KNO
distributions $\Psi(z)$ in the domain $1<\etas<5$ for four intervals in
$W$, plotted on a logarithmic (top) and a linear scale (bottom),
respectively. The dotted curve is the KNO function for $e^+e^-$
annihilation\footnote{Single hemisphere $e^+e^-$ data are known to exhibit
KNO scaling above $\approx 20$ GeV~\cite{tasso:mul,opal:mul,delphi:md}.
The JETSET simulations show the same in the domain $1<\etas<5$ discussed
here.} in the same $\etas$ range, obtained from the JETSET parton shower
model at a centre-of-mass energy of 91.2 GeV. The two data sets are
remarkably similar.

Exact KNO scaling implies that the function $\Psi(z)$, and hence the
moments $C_q$, are independent of $W$. Fig.~6 displays the variation of
$C$ moments with $W$ for various $\etas$ intervals (see also Table~4).
At fixed $W$ the moments increase as the $\etas$ domain decreases in
size, reflecting the widening of the multiplicity distribution in
KNO form.

The moments in the smallest and largest pseudorapidity domain show,
within errors, little $W$ dependence and thus exhibit approximate 
KNO scaling. However,  violation of KNO scaling is
seen in intermediate size intervals.
This observation is consistent with the clear KNO scaling
violations observed at HERA in DIS data on multiplicity distributions
measured in the current region of the Breit frame of 
reference~\cite{zeus:md,h1:thompson}. 

The MEPS~6.4 generator  (solid line) describes the data well in the
largest $\etas$ domains but tends to underestimate $C_2$ and $C_3$ in
the smaller ones. %For $1<\etas<5$ it exhibits approximate KNO scaling.

%In Fig.~6 we further show TASSO~\cite{tasso:mul} and
%DELPHI~\cite{delphi:md} data on $C$ moments measured, however, in the
%centre-of-mass {\em rapidity} interval $0<y<y_c$ ($y_c=1,2,3,4$).
%It is seen that the absolute values of the moments, and the
%evolution with energy, are  comparable to those in DIS data in large
%domains. However, in the smallest  $y$ domain, they  differ quite strongly.

The dotted lines in Fig.~6 are expectations from JETSET for $e^+e^-$ in the
same pseudorapidity domains as  covered by  H1. Large differences
between DIS and $e^+e^-$ annihilation are predicted, in particular for
higher order moments in small $\etas$ domains. However, in the largest
domains the $e^+e^-$ results join smoothly with the DIS data. It will be of
interest to confirm  these predictions with future measurements at LEP in
the $W$ range studied here.

Also shown is a measurement for non-single-diffractive \pp collisions at
$\sqrt{s}=200$~GeV\footnote{The UA5 measurement at $\sqrt{s}=900$~GeV
(not shown) yields $C_2=1.84\pm0.02$, $C_3=4.0\pm0.1$ and $C_4=14.5\pm0.8$,
the same, within errors, as the results at $200$ GeV.} in the interval
$1<|\etas|<2$  from UA5~\cite{ua5:md:2:9}. Here the fluctuations in
particle density number near the central plateau are significantly
larger than in DIS.  It will, however, be shown in Sect.~\ref{sec:et} that
the particle density itself is quite similar in the two processes.

The values of the cumulants $K_q$ in a given domain of phase space are a
direct measure of the strength of ``genuine'' correlations
among hadrons. Inspection of Table~4 shows
that the three-particle correlation function is significantly different
from zero only in the smallest  interval $1<\eta^\ast<2$, in accord with
measurements for other types of interactions~\cite{review:93b}.
 
To study the two-particle correlation function more directly we present
in Fig.~7 $\avgn$  and the second order factorial moment $R_2=1+K_2$ 
in $\etas$ domains. The  mean charged multiplicity
increases approximately as $\ln W$ for $1<\etas<2$, i.e.~near the
central region, but faster in larger domains. 
The $W$ dependence of $R_2$ for $1<\etas<2$  is less clear,
in view of the errors, but compatible with the slow (logarithmic) rise 
well established  in hadron-hadron   and $e^+e^-$ interactions. The
MEPS~6.4 generator  reproduces reasonably well the behaviour of $R_2$
but  systematically overestimates the mean
multiplicity  in all  $\eta^\ast$ domains.

%=======================================================================

\subsubsection{Particle density and $E_T$ flow\label{sec:et}}

The flow of transverse energy, $E_T$, in multiparticle final states at
high energy is studied intensively at hadron colliders and by the HERA
experiments. The $E_T$ distribution in a given phase space domain, and
its moments, are convoluted observables. They depend not only on the
particle density and the $p_T$ structure of  the collisions but also on
multiparticle correlations, and, therefore, on the moments of the
multiplicity distribution in that domain~\cite{weiner:et}. Here we
compare the evolution with $W$ of the particle density to that of the
mean $E_T$ for DIS and hadron-hadron interactions.

Fig.~8 shows a compilation of measurements in
DIS~\cite{e665:mul,emc:mul} and in non-single-diffractive hadron-hadron
collisions~\cite{ua5:md:2:9,na22:neg:binom1,adr:thesis,sfm:mul,ua1:mul}
of mean charged  multiplicity per unit of rapidity or pseudorapidity
(solid symbols). The H1 data are the same as those shown in Fig.~7a. The
open symbols show measurements of the mean $E_T$ in the region
$-0.5<\etas<0.5$, presented and discussed in~\cite{h1:dis:photo}. The
solid line shows a parameterisation used by UA1~\cite{ua1:mul} with the
form $\avgn=0.35+0.74\,(W^2)^{0.105}$.

In spite of the differences in the rapidity region covered, 
the known difference
between pseudorapidity and rapidity density, different experimental
procedures and systematics, it remains of interest to note that the
charged particle density for DIS at HERA interpolates quite smoothly
with DIS and hadron-hadron data at lower and much higher energy. This
observation is consistent with the analogy between
virtual photon-hadron, real photon-hadron  and hadron-hadron
interactions originally advocated  by Gribov and Feynman\footnote{For
further discussion and references see~\cite{doksh:91}, Chapt.~4.}, 
which suggests universality of dynamics in the
central plateau~\cite{h1:dis:photo}. 
However, the difference in correlation
strength, noted in the previous section, suggests that such
``universality'', while applicable to single-inclusive
spectra~\cite{h1:dis:photo}, may not hold for higher order correlations.

Several mechanisms have been suggested to explain the dynamics of the
large multiplicity fluctuations in soft hadron-hadron collisions:
impact parameter averaged  Poisson-like fluctuations; multiple soft
parton interactions in the same event leading to
mini-jets~\cite{pythia}; multi-pomeron exchange as in the Dual Parton
Model~\cite{capella:dpm} and multi-string configurations as in the Lund
FRITIOF model~\cite{fritiof}. These mechanisms have no direct analogues
in DIS, except for the boson-gluon fusion QCD process which can lead to
two-string colour topologies and could, therefore, mimic multi-string
properties in hadron-hadron models.

In Fig.~8 we further compare the energy evolution of $\avgn$ in the
central region in DIS with that in $e^+e^-$ annihilation. The dotted curve
is  the prediction  from  the JETSET generator for the interval 
$1<\etas<2$. 
It shows  that the hadron density  evolves much faster with $W$ 
than in the DIS and hadron-hadron data.

The energy evolution of $\avgn$ in perturbative QCD is controlled by
the anomalous multiplicity dimension $\gamma$ through $\gamma = {\rm d}
\ln \avgn / {\rm d} \ln W^2$. For $e^+e^-$ annihilation, and in restricted
(pseudo)rapidity intervals, it is given by
$\gamma_{ee}(\eta^\ast)\sim\sqrt{3\alpha_s(k^2_{Tmax})/2\pi}$, where
$k_{Tmax}$ is the maximum possible transverse momentum at a given
$\eta^\ast$~\cite{lund:anom}. From the JETSET $e^+e^-$ predictions above
$W=20$ GeV we derive that $\gamma_{ee}$ is constant with a value of
$0.16$. This is somewhat smaller than expected from the analytic result
($\sim0.22$ at 200 GeV) but agrees with that quoted in~\cite{lund:anom}.

In leptoproduction, where gluon emission from the nucleon remnant is
thought to be suppressed, $\gamma_{DIS}(\etas)$ depends more
strongly on $\etas$ with
$\gamma_{DIS}(\etas)\sim(1/2)\gamma_{ee}(\etas)$ for not too
large  positive $\etas$~\cite{lund:anom}. 
From the DIS data in Fig.~8, using the UA1
parameterisation, we  estimate $\gamma_{DIS}=0.8-0.9$, consistent with
expectations. This is the first semi-quantitative experimental
confirmation of the ``antenna suppression'' effect in DIS. However, better data
are needed,  also at lower $W$, to exploit these perturbative QCD
predictions in a fully quantitative manner.
This result also implies that the rate of increase  with energy of 
$\aver{n}$ in the full current hemisphere for DIS and that in a single
hemisphere for $e^+e^-$ annihilations, discussed in Sect.~4.1.1,
are expected to differ at higher centre of mass energies.

Non-asymptotic analytical QCD predictions for higher order multiplicity
moments are at present not available. It is known, however, that the
energy dependence of the moments is asymptotically 
controlled by $\gamma_{DIS}(\etas)$
or  $\gamma_{ee}(\etas)$~\cite{Dahl89}. It is therefore likely that the
differences between the $C$ moments in DIS and $e^+e^-$, described in
Sect.~\ref{sec:kno}, are a further  reflection of suppressed gluon
emission in DIS.

Finally, the comparison between $\avgn$ and $\aver{E_T}$ in Fig.~8
demonstrates that both evolve with energy rather similarly. This suggests
that the increase of mean transverse energy with increasing $W$
(decreasing $x_B$) at fixed $Q^2$, previously observed in this
experiment~\cite{h1:etflow:95}, follows mainly from an increase of the
hadron multiplicity and less so from a rise of the mean transverse
momentum of individual hadrons. A direct measurement of the energy
dependence of the transverse momentum distribution in or near the central
region of deep inelastic $ep$ collisions at HERA, and comparison with
existing hadron-hadron data, should help to clarify this interesting and
theoretically much debated issue. 

%=======================================================================

\section{Summary}

%In this paper we presented  the first results
%on the  multiplicity distribution of charged hadrons in the current
%fragmentation region of the hadronic centre-of-mass in
% deep inelastic scattering  $e^+p$ collisions at HERA,
%using the H1 detector.
%The analysis used  data accumulated in 1994,  corresponding to
%an integrated luminosity of $1.3$~pb${}^{-1}$.

Data, fully corrected for detector effects, are presented on the evolution
with $W$ and $Q^2$ of the charged particle multiplicity distribution and
its statistical moments, over the ranges $80<W<220$ GeV and $10<Q^2<1000$
GeV${}^2$ in subdomains of pseudorapidity space, including the full
current hemisphere. The main results can be summarised as follows:

\begin{itemize}

\item The mean charged hadron multiplicity and the dispersion, measured in
fixed intervals of $W$  and in the  domain $1<\etas<5$ show,
within errors, no dependence on the virtuality of the exchanged boson over
the $Q^2$ range covered by H1.

\item The low order moments of the multiplicity distribution in the full
current hemisphere show noteworthy similarities with single hemisphere
data in $e^+e^-$ annihilation, in conformity with the hypothesis of approximate
environmental independence of quark hadronisation.  In particular, the mean
charged hadron multiplicity in DIS shows a similar rate of increase with
$W$ to that measured in $e^+e^-$ annihilation up to $130$ GeV. Analytical
predictions from perturbative QCD on the mean parton multiplicity in jets,
which are proven to be valid for hadrons in $e^+e^-$ annihilation, are
therefore confirmed for the first time in DIS at higher energies than
presently reachable at LEP.  Data on the second order factorial moment
show that higher order QCD corrections and/or non-perturbative effects
remain significant at present energies.

\item The analysis of the \MD{} in pseudorapidity domains of varying size
proves that the well-documented property of KNO scaling, a general
characteristic of scale-invariant stochastic branching processes, remains
valid in DIS at HERA for small and for large pseudorapidity intervals, but
not in intermediate size domains. The KNO phenomenon, also predicted in
QCD at asymptotic energies, results from an intricate interplay of
correlations, different in different types of collisions, and changing
rapidly over phase space.  The KNO function in the region $1<\eta^\ast<5$
is strikingly similar to that expected for $e^+e^-$ annihilation under the
same kinematical conditions. 

\item The charged particle density near the central region of the
$\gamma^* p$ centre-of-mass system grows significantly more slowly in DIS
than in $e^+e^-$ annihilation.  The strength of particle correlations, as
reflected in the $C$ moments, is much larger in the latter 
process for small pseudorapidity domains. Such differences can be
understood within perturbative QCD from calculations of the
local anomalous multiplicity dimensions within the Lund dipole formalism. 
The comparison between $e^+e^-$ annihilation and DIS data provides direct
evidence for the ``antenna suppression'' effect in current fragmentation,
a characteristic of deep inelastic lepton-hadron dynamics at small
Bjorken-$x$.
The same mechanism offers a qualitative explanation for the absence of a
multi-jet induced shoulder structure in the DIS multiplicity distributions
in intermediate size pseudorapidity domains, now well established in
$e^+e^-$ annihilation data at LEP. 

%In the largest pseudorapidity domains investigated the correlations, as
%well as the mean multiplicity, are essentially of the same magnitude. 
%This illustrates that differential measurements are needed to reveal
%differences in dynamics. 

\item The charged particle density near the central plateau in the
deep inelastic process is of the same magnitude as that for minimum bias
non-single-diffractive hadron-hadron interactions at the same value of the
centre-of-mass energy.  Its evolution with $W$ is also comparable to that
in hadron-hadron collisions measured up to $900$ GeV.  However,
hadron-hadron collisions are characterised by substantially stronger
correlations in small pseudorapidity domains.  These features remain to be
understood within the Gribov-Feynman pictures of DIS.

\item A comparison of the evolution with $W$ of mean transverse energy
flow and charged particle density near the central region in DIS shows
that the two phenomena are strongly correlated.  The striking similarity
with the behaviour seen in non-single-diffractive hadron-hadron collisions
implies that a theoretical explanation within QCD should simultaneously
address the dynamics of both types of process. 

\item The multiplicity distributions in the smallest $\etas$ domain
examined can be well parameterised with Lognormal or Negative Binomial
functions.  However, the quality of the fits deteriorates in larger
domains.  The large $n$ tail of the experimental distributions is well
described but deviations occur for small multiplicities.  Nevertheless, as
economic representations of the data, these parametric forms continue to
be useful for phenomenology.

\item Among a variety of Monte Carlo generators presently being developed
for DIS, we have shown predictions from MEPS~6.4.  This model
overestimates the mean charged multiplicity in the full current
hemisphere, as well as in subdomains of pseudorapidity space. Higher order
moments of the multiplicity distributions are reasonably well described. 

%Similar conclusions are reached for other generators which have been
%investigated in the course of this study.

\end{itemize}

%The results presented in this paper show that differential measurements of
%hadron correlations are able to reveal differences in the momentum space
%structure or geometry~\cite{bj:vietri} of multiparticle final states
%produced in different types of collision process. In future work, with
%the help of recently developed refined analysis
%techniques~\cite{review:93b}, it should become possible to extract from
%present and future experiments sufficiently discriminative information to
%test various competing theoretical approaches to the dynamics of
%deep inelastic scattering at very small $x_B$. 

%=======================================================================

\vspace*{1.cm}          

\noindent {\bf Acknowledgements}  
\normalsize    
        
\noindent We are very grateful to the HERA machine group whose outstanding
efforts made this experiment possible. We acknowledge the support of the
DESY technical staff.  We appreciate the big effort of the engineers and
technicians who constructed and maintain the detector. We thank the
funding agencies for financial support of this experiment.  We wish to
thank the DESY directorate for the support and hospitality extended to the
non-DESY members of the collaboration. 
Finally, we thank I.~Dremin, G.~Gustafson, V.~Khoze  and W.~Ochs for
stimulating discussions.        
%=======================================================================
       
\vspace*{.2cm}
\newpage
\bibliography{multi}

%=======================================================================

\newpage

\input newtab.tex

%=======================================================================

\begin{table}[htbp]
\begin{center}
{\scriptsize
\begin{tabular}
{ 
c 
r@{.}l @{ $\pm$ } r@{.}l 
r@{.}l @{ $\pm$ } r@{.}l
r@{.}l @{ $\pm$ } r@{.}l
r@{.}l @{ $\pm$ } r@{.}l }
\multicolumn{17}{c}{{\small $1 < \eta^\ast < 2$}} 
\\ \\
$W$ (GeV) &
\multicolumn{4}{c}{$80 \div 115$} &
\multicolumn{4}{c}{$115 \div 150$} &
\multicolumn{4}{c}{$150 \div 185$} &
\multicolumn{4}{c}{$185 \div 220$} \\
 \hline \\
$\nbar$ & 2&52  & 0&10  & 2&58  & 0&13  & 2&72  & 0&15  & 2&75  & 0&19  \\
$1/k$ & 0&285 & 0&080 & 0&288 & 0&056 & 0&269 & 0&034 & 0&270 & 0&064 \\
\chnd & \chndf{29.4}{13} & \chndf{23.0}{14} & \chndf{33.0}{15} & \chndf{27.1}{14} \\
\\
$m$   & 2&44  & 0&16  & 2&49  & 0&16  & 2&66  & 0&20  & 2&67  & 0&13  \\
$d$   & 0&791 & 0&092 & 0&812 & 0&077 & 0&743 & 0&094 & 0&768 & 0&122 \\
$c$   & 2&6   & 3&4   & 3&1   & 1&2   & 1&9   & 1&7   & 2&5   & 7&5 \\
\chnd & \chndf{21.8}{12} & \chndf{8.9}{13} & \chndf{24.9}{14} & \chndf{19.8}{13} \\
\\
\\
\multicolumn{17}{c}{{\small $1 < \eta^\ast < 3$}} 
\\ \\
$W$ (GeV) &
\multicolumn{4}{c}{$80 \div 115$} & 
\multicolumn{4}{c}{$115 \div 150$} &
\multicolumn{4}{c}{$150 \div 185$} & 
\multicolumn{4}{c}{$185 \div 220$} \\
\hline \\
$\nbar$ & 4&85  & 0&16  & 5&12  & 0&26  & 5&36  & 0&25  & 5&41  & 0&20  \\
$1/k$ & 0&161 & 0&039 & 0&235 & 0&017 & 0&256 & 0&028 & 0&245 & 0&034 \\
\chnd & \chndf{121.6}{19} & \chndf{54.5}{19} & \chndf{61.5}{21} & \chndf{44.7}{24} \\
\\
$m$   & 4&79  & 0&24  & 5&05  & 0&30  & 5&25  & 0&20  & 5&32  & 0&17  \\
$d$   & 0&634 & 0&017 & 0&644 & 0&080 & 0&662 & 0&074 & 0&641 & 0&060 \\
$c$   & 4&5   & 2&2   & 2&0   & 6&5   & 2&1   & 5&5   & 1&8   & 6&3 \\
\chnd & \chndf{71.1}{18} & \chndf{42.7}{18} & \chndf{43.6}{20} & \chndf{31.1}{23} \\
\\
\\
\multicolumn{17}{c}{{\small $1 < \eta^\ast < 4$}}
\\ \\
$W$ (GeV) &
\multicolumn{4}{c}{$80 \div 115$} &
\multicolumn{4}{c}{$115 \div 150$} &
\multicolumn{4}{c}{$150 \div 185$} &
\multicolumn{4}{c}{$185 \div 220$} \\
\hline \\
$\nbar$ & 6&41  & 0&28  & 7&00  & 0&39  & 7&52  & 0&46  & 7&71  & 0&30  \\
$1/k$ & 0&085 & 0&018 & 0&102 & 0&014 & 0&132 & 0&026 & 0&142 & 0&020 \\
\chnd & \chndf{96.8}{21} & \chndf{112.4}{21} & \chndf{88.6}{25} & \chndf{44.7}{24} \\
\\
$m$   & 6&40  & 0&33  & 6&98  & 0&37  & 7&45  & 0&41  & 7&59  & 0&28  \\
$d$   & 0&475 & 0&009 & 0&484 & 0&011 & 0&505 & 0&023 & 0&527 & 0&022 \\
$c$   & 2&7   & 1&3   & 2&7   & 2&5   & 2&8   & 7&1   & 4&5   & 5&9 \\
\chnd & \chndf{62.6}{20} & \chndf{69.6}{20} & \chndf{43.8}{24} & \chndf{32.8}{23} \\
\\
\\
\multicolumn{17}{c}{{\small $1 < \eta^\ast < 5$}}
\\ \\
$W$ (GeV) &
\multicolumn{4}{c}{$80 \div 115$} &
\multicolumn{4}{c}{$115 \div 150$} &
\multicolumn{4}{c}{$150 \div 185$} &
\multicolumn{4}{c}{$185 \div 220$} \\
\hline \\
$\nbar$ & 6&90  & 0&33  & 7&73  & 0&41  & 8&44  & 0&49  & 8&88  & 0&34  \\
$1/k$ & 0&067 & 0&011 & 0&068 & 0&012 & 0&074 & 0&013 & 0&069 & 0&020 \\
\chnd & \chndf{44.7}{19} & \chndf{57.0}{21} & \chndf{80.8}{24} & \chndf{63.6}{25} \\
\\
$m$   & 6&89  & 0&29  & 7&72  & 0&41  & 8&40  & 0&48  & 8&78  & 0&33  \\
$d$   & 0&433 & 0&010 & 0&422 & 0&007 & 0&422 & 0&015 & 0&422 & 0&020 \\
$c$   & 1&7   & 1&5   & 1&8   & 2&0   & 2&4   & 5&1   & 4&1   & 6&1 \\
\chnd & \chndf{41.6}{18} & \chndf{52.8}{20} & \chndf{58.4}{23} & \chndf{27.7}{24} \\
\end{tabular}}
\end{center}
\label{tab:fitpar}
\caption{Parameters of  Negative Binomial and Lognormal 
distributions in  domains of pseudorapidity $\eta^\ast$ and $W$. 
Errors quoted are the quadratic sum of the statistical error
and systematic uncertainties.}
\label{tab:moments_in_rapidity_ranges}
\end{table}

\begin{figure}[htbp]
\epsfig{figure=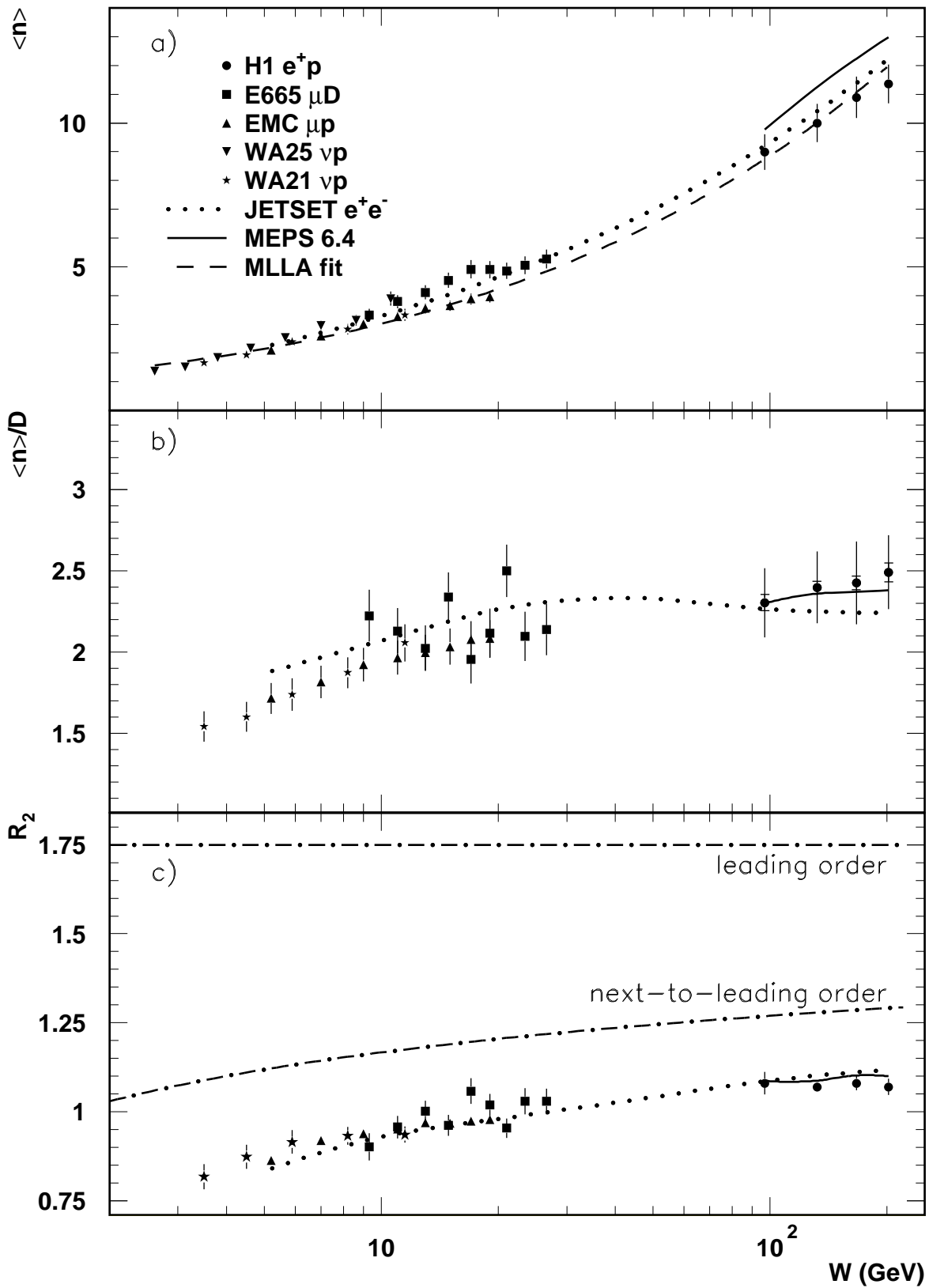}
\label{fig:fullhemi}
\caption{ {\bf(a)} $W$ dependence of the mean charged particle multiplicity 
in the full current hemisphere in comparison with 
fixed target lepton-nucleon data. The various   curves are described in the
text; {\bf(b)} and {\bf(c)}  same as {\bf(a)} for the ratio 
$\langle n \rangle/D$ and the second 
factorial moment $R_2$, respectively.} 
\end{figure}

\begin{figure}[htbp]
\epsfig{figure=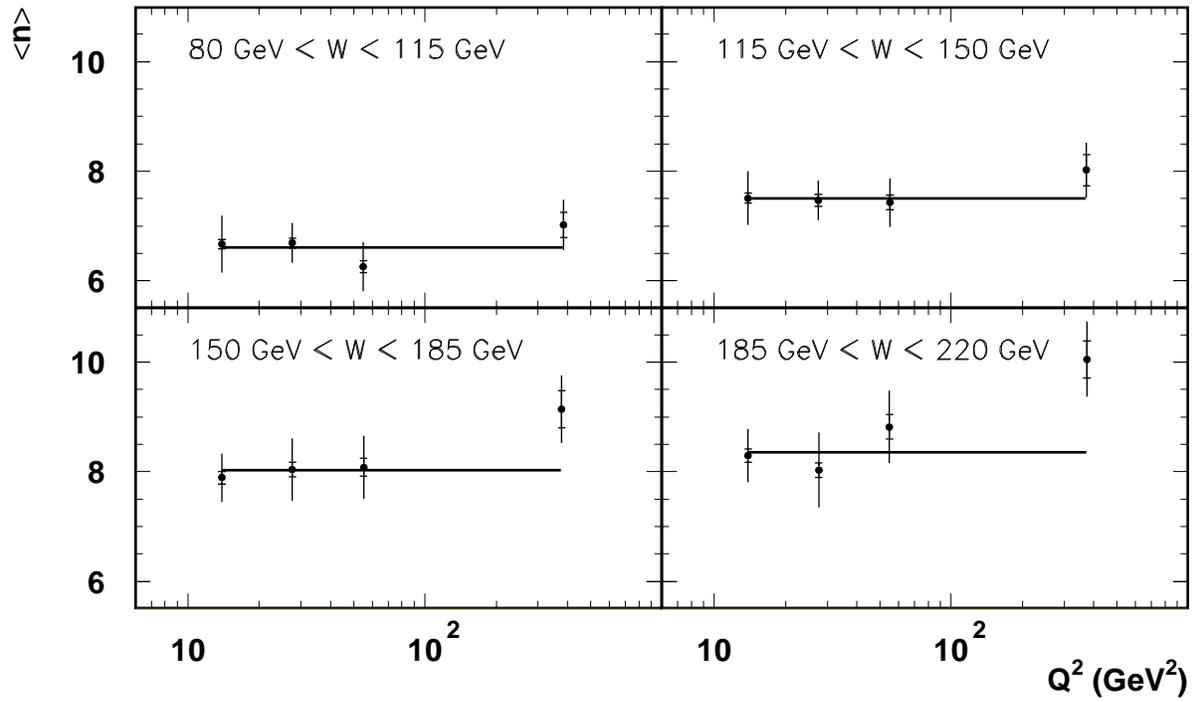}
\epsfig{figure=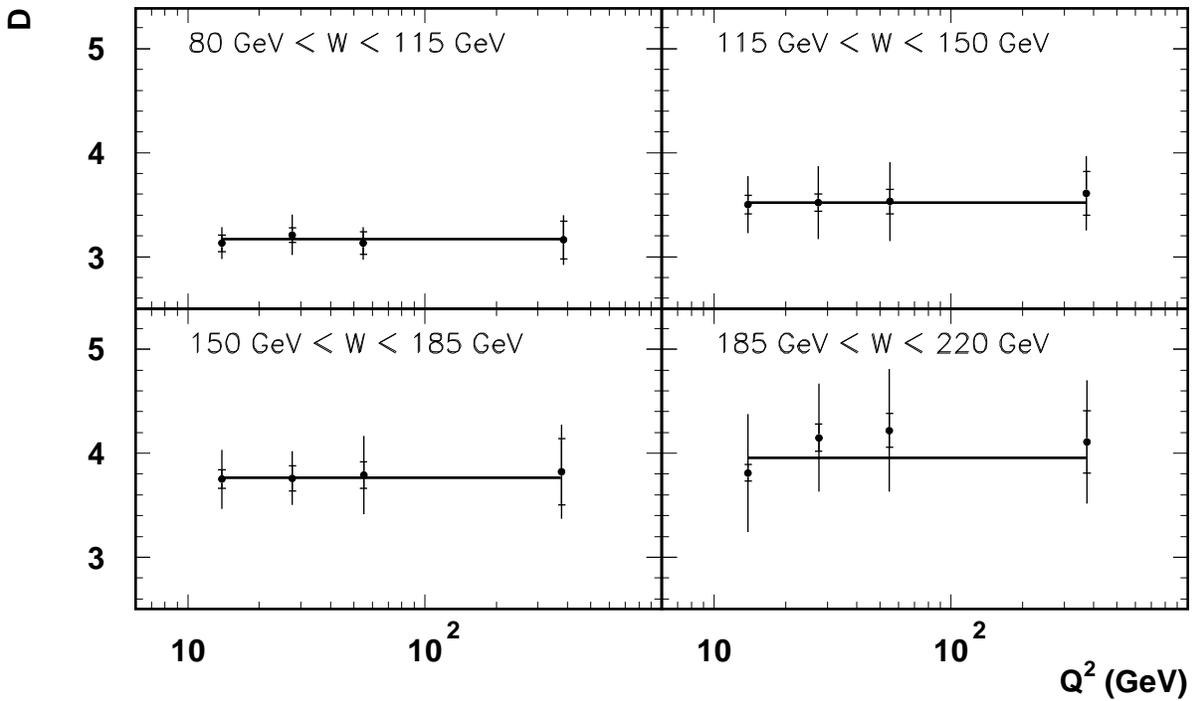}
\label{fig:q2dep}
\caption{The $Q^2$ dependence of the mean charged particle multiplicity 
(top) and the dispersion (bottom) in intervals of $W$ and
in the domain $1<\etas<5$. Horizontal lines are fits to a constant. The 
inner error bars represent the statistical error, the outer error bars 
represent the total (quadratic sum of statistical and systematical) errors.} \end{figure}

\begin{figure}[htbp]
\epsfig{figure=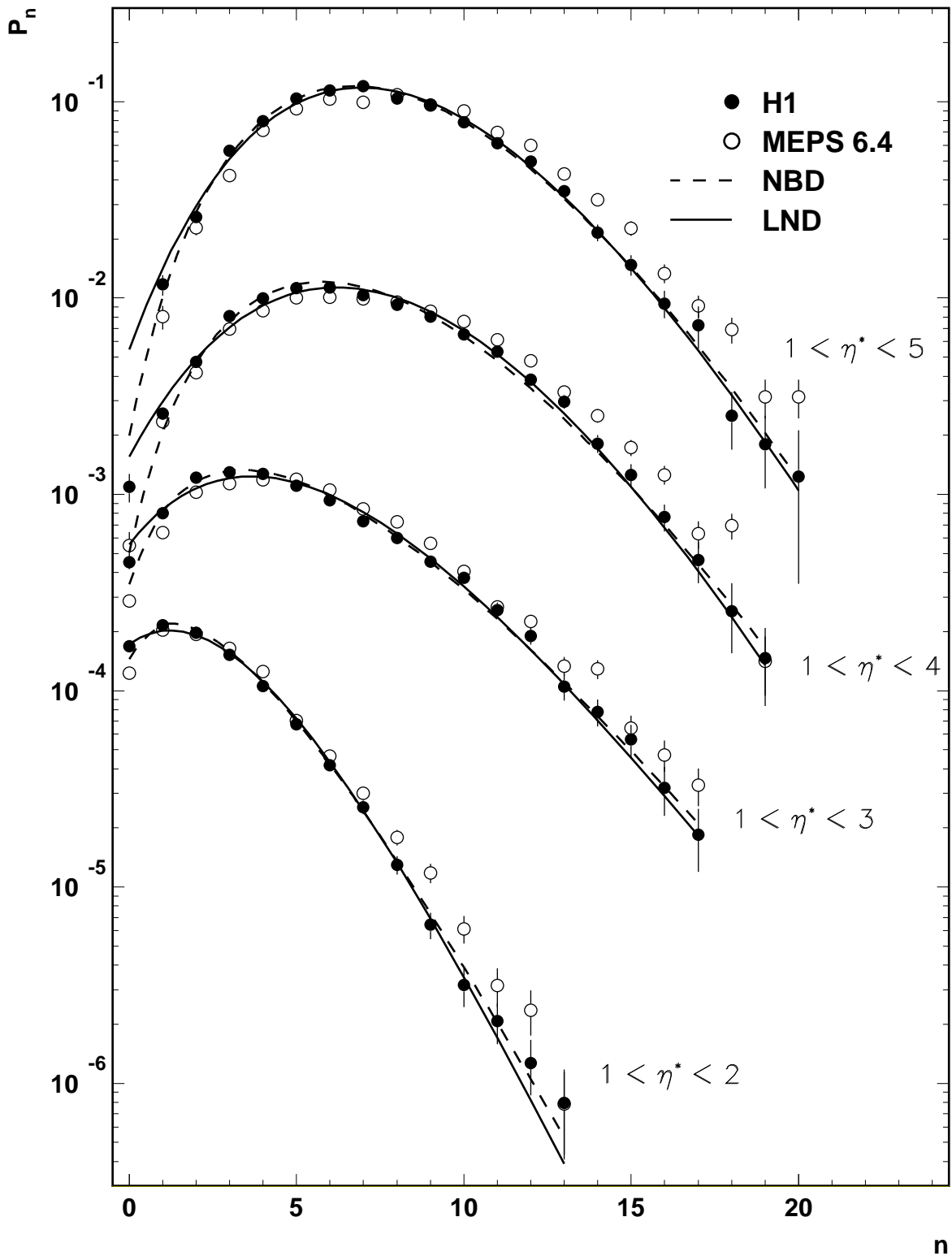}
\label{fig:fits}
\caption{The unfolded multiplicity distribution in the interval 115 GeV 
$< W <$ 150 GeV  and $Q^2>10$~GeV${}^2$, in indicated pseudorapidity domains. 
The 
distribution for $1 < \eta^\ast < 5$ is plotted at its true scale; each 
consecutive distribution is shifted down by factor  of 10. The H1 data points 
(solid symbols) are compared to  MEPS 6.4  (open symbols) and to 
fits with the Lognormal (full) and Negative Binomial (dashed) distribution.
Statistical errors only are plotted.}  
\end{figure}

\begin{figure}[htbp]
\epsfig{figure=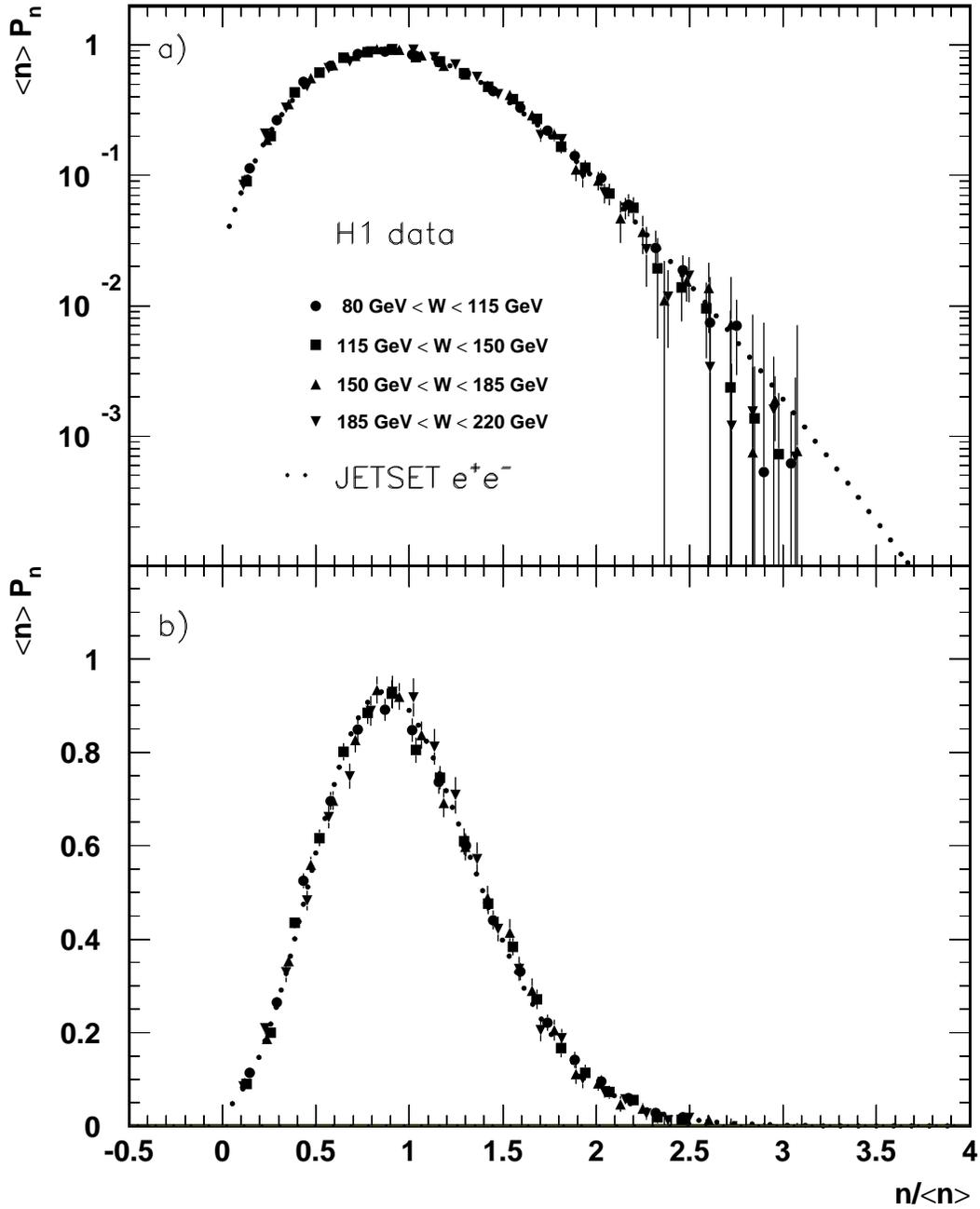}
\caption{Unfolded multiplicity distributions in the pseudorapidity 
domain $1 < \eta^\ast < 5$ measured in different $W$ intervals plotted in 
KNO form, in logarithmic and linear scale. Statistical errors only are 
plotted.} \end{figure}

\begin{figure}[htbp]
\epsfig{figure=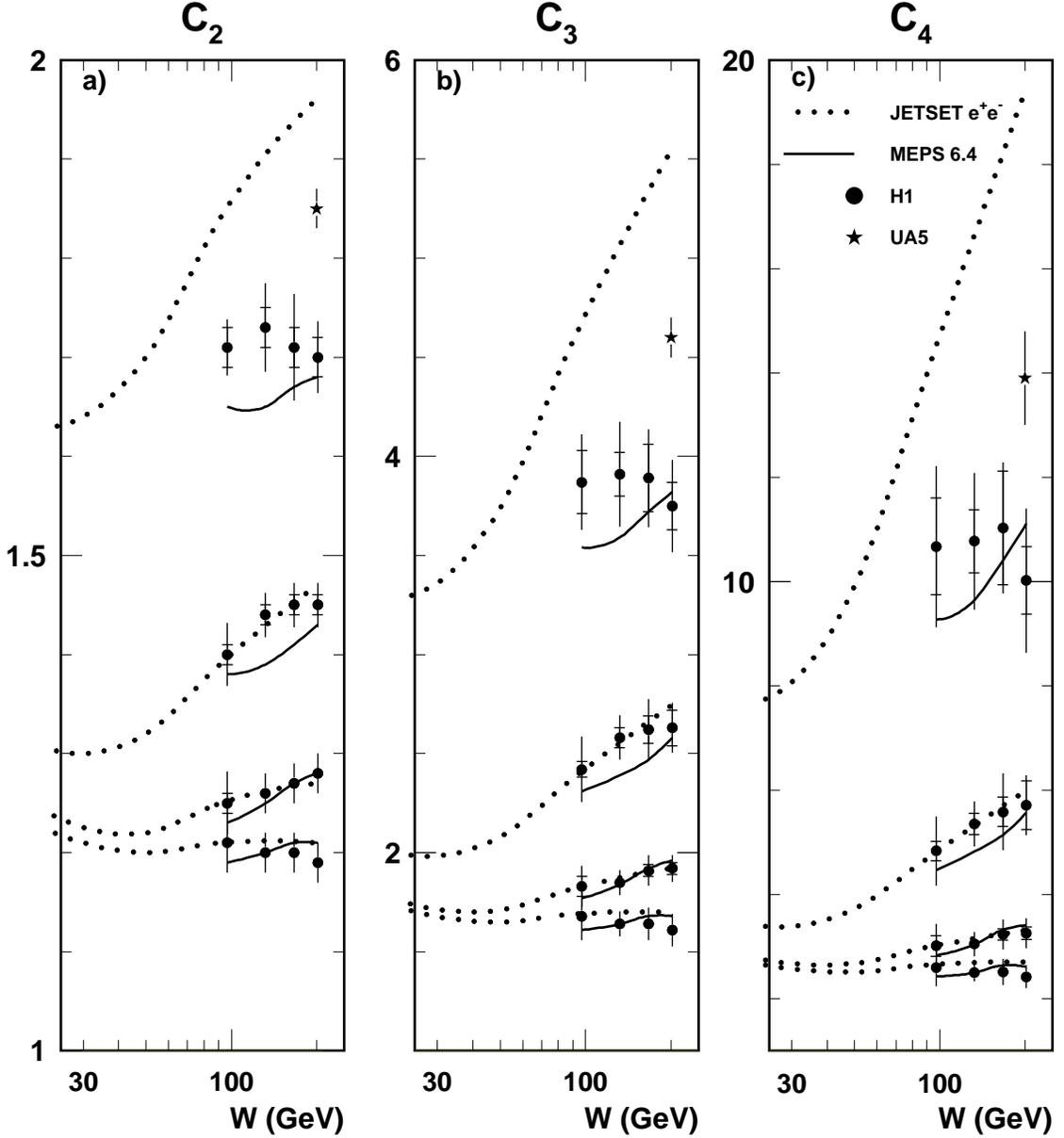}
\caption{$W$ dependence of the multiplicity moments $C_q$ in various 
$\eta^\ast$ domains. Data from UA5~\protect\cite{ua5:md:2:9} in the
interval $1<|\eta^\ast|<2$ are also shown.
  From top to  bottom, the domains are 
$1 < \eta^\ast < 2$,  
$1 < \eta^\ast < 3$, 
$1 < \eta^\ast < 4$ 
and $1 < \eta^\ast < 5$. The curves are described in the text.
The inner error bars represent the statistical errors, the outer error 
bars represent the total (quadratic sum of statistical and systematical) errors.} 
\end{figure}

\begin{figure}[htbp]
\epsfig{figure=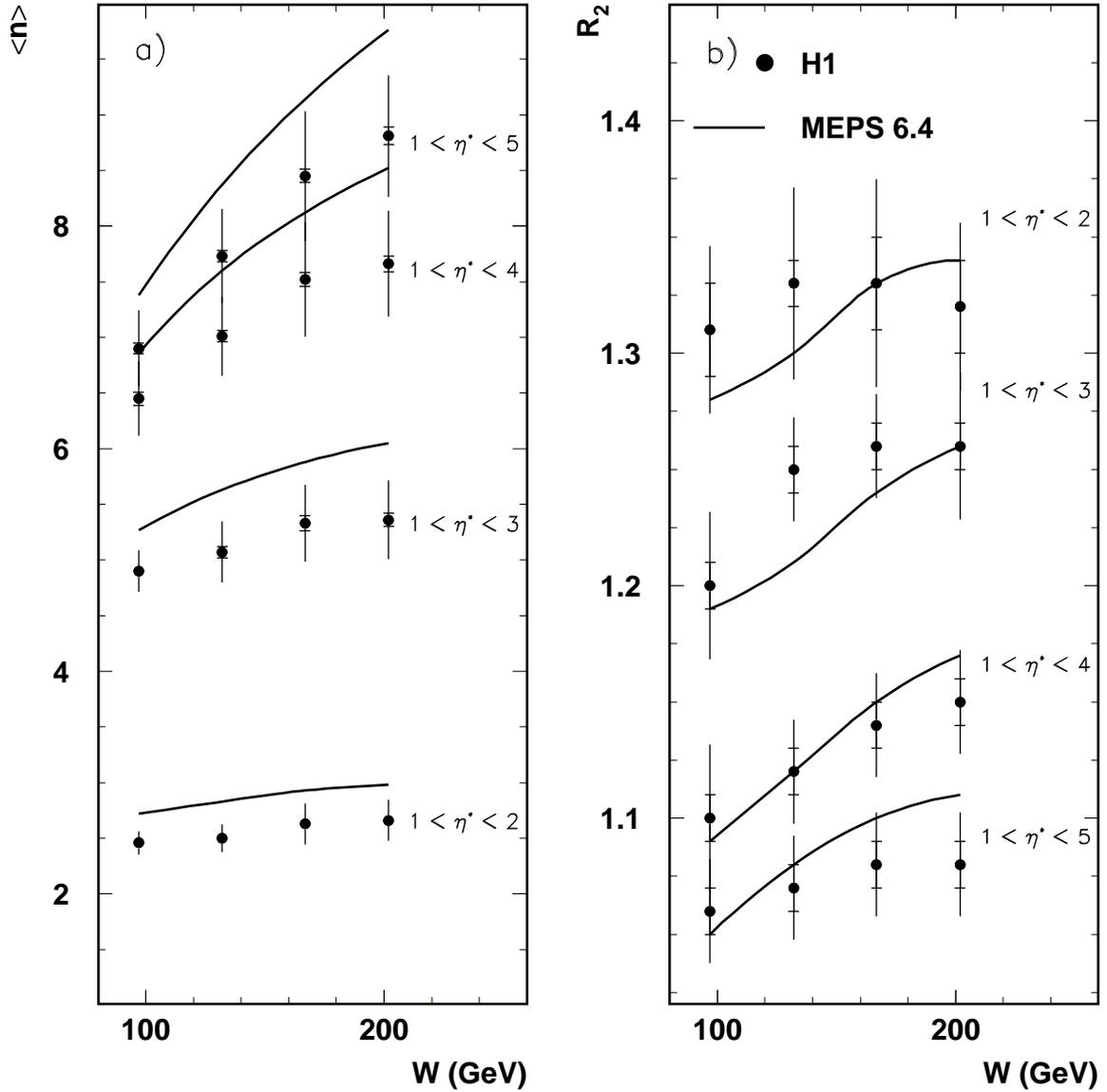}
\caption{$W$ dependence of mean charged multiplicity and
second order factorial moment $R_2$ in indicated
pseudorapidity domains, compared with MEPS~6.4 predictions. The inner error bars represent the statistical errors, the outer error 
bars represent the total (quadratic sum of statistical and systematical) errors.}
\end{figure}

\begin{figure}[htbp]
\epsfig{figure=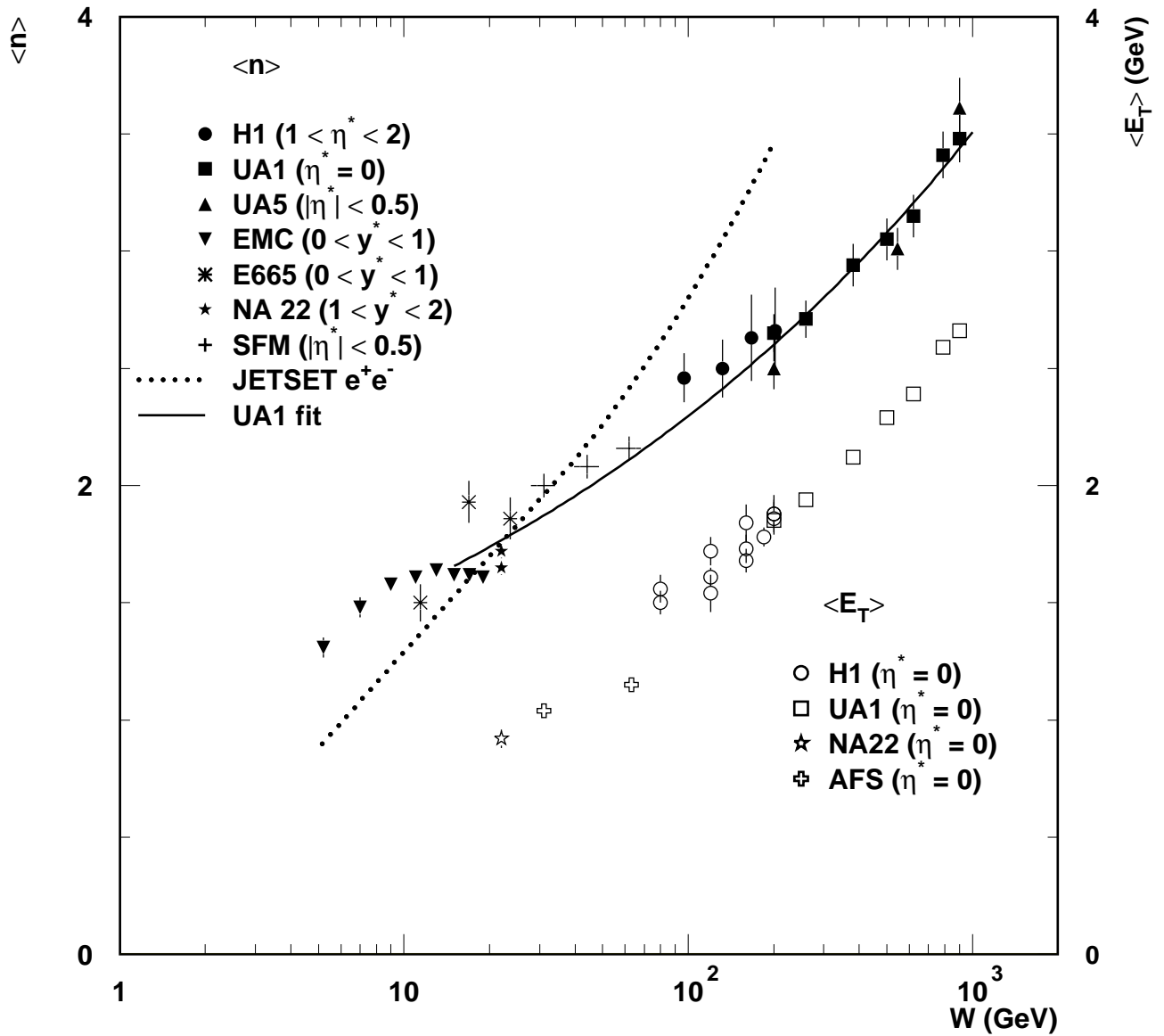}
\caption{Solid symbols: $W$ dependence of charged particle
density in DIS and hadron-hadron collisions (left scale); open symbols:
the mean transverse momentum flow, per unit of (pseudo)rapidity (right
scale).
The curves are described in the text. $\etas$ is pseudorapidity, while 
$y^\ast$ is rapidity in the hadronic centre-of-mass frame.} 
\end{figure}

\end{document}

%% file: newtab.tex
\begin{table}[p]
\begin{center}
{\scriptsize
\begin{tabular}{r@{\ }c@{\ }r r@{.}l r@{\ $\ra$\ }r r@{.}l r }
\multicolumn{3}{c}{$W$ range}  &
\multicolumn{2}{c}{$\langle W \rangle$} &
\multicolumn{2}{c}{$Q^2$ range} &
\multicolumn{2}{c}{$\langle Q^2 \rangle$} &
$\sharp$ events \\
\multicolumn{3}{c}{(GeV)} &
\multicolumn{2}{c}{(GeV)} &
\multicolumn{2}{c}{(\gevsq)} &
\multicolumn{2}{c}{(\gevsq)} &
\\ \hline \\
 80 & $\ra$ & 115 &  96&9                 &  10 & 20   &  13&9 &  9150 \\
    &       &     & \multicolumn{2}{c}{ } &  20 & 40   &  27&6 &  5021 \\
    &       &     & \multicolumn{2}{c}{ } &  40 & 80   &  55&0 &  2509 \\
    &       &     & \multicolumn{2}{c}{ } & 200 & 1000 & 385&3 &   377 \\
    &       &     & \multicolumn{2}{c}{ } &  10 & 80   &  22&9 & 16680 \\ \\
115 & $\ra$ & 150 & 132&0                 &  10 & 20   &  13&9 &  8202 \\
    &       &     & \multicolumn{2}{c}{ } &  20 & 40   &  27&5 &  4360 \\
    &       &     & \multicolumn{2}{c}{ } &  40 & 80   &  55&1 &  2421 \\
    &       &     & \multicolumn{2}{c}{ } & 200 & 1000 & 372&8 &   411 \\
    &       &     & \multicolumn{2}{c}{ } &  10 & 80   &  23&2 & 14983 \\ \\
150 & $\ra$ & 185 & 166&8                 &  10 & 20   &  13&9 &  6778 \\
    &       &     & \multicolumn{2}{c}{ } &  20 & 40   &  27&6 &  3662 \\
    &       &     & \multicolumn{2}{c}{ } &  40 & 80   &  55&1 &  1751 \\
    &       &     & \multicolumn{2}{c}{ } & 200 & 1000 & 378&4 &   439 \\
    &       &     & \multicolumn{2}{c}{ } &  10 & 80   &  23&3 & 12191 \\ \\
185 & $\ra$ & 220 & 201&9                 &  10 & 20   &  13&9 &  5299 \\
    &       &     & \multicolumn{2}{c}{ } &  20 & 40   &  27&6 &  2919 \\
    &       &     & \multicolumn{2}{c}{ } &  40 & 80   &  54&9 &  1037 \\
    &       &     & \multicolumn{2}{c}{ } & 200 & 1000 & 374&2 &   349 \\
    &       &     & \multicolumn{2}{c}{ } &  10 & 80   &  23&5 &  9255 \\
\end{tabular}}
    
\caption{Number of events, mean $Q^2$ and mean $W$
in the different kinematical regions studied.}
\label{binstat}
\end{center}
\end{table}

\begin{table}[p]
\begin{center}
{\scriptsize
\begin{tabular}{p{1.7cm} r@{.}l  r@{.}l  r@{.}l  r@{.}l | r@{.}l  r@{.}l  
r@{.}l
r@{.} l}
  & \multicolumn{8}{c|}{\small $\eta^\ast>0$}
  & \multicolumn{8}{c}{\small $1 < \eta^\ast < 3$} \\
$W$ (GeV)
  & \multicolumn{2}{c}{80 $\div$ 115} & \multicolumn{2}{c}{115 $\div$ 150}
  & \multicolumn{2}{c}{150 $\div$ 185} & \multicolumn{2}{c|}{185 $\div$ 220}
  & \multicolumn{2}{c}{80 $\div$ 115} & \multicolumn{2}{c}{115 $\div$ 150}
  & \multicolumn{2}{c}{150 $\div$ 185} & \multicolumn{2}{c}{185 $\div$ 
220}\\ \hline
\multicolumn{9}{c|}{ } & \multicolumn{8}{c}{ } \\
photo-       & $-$0&7\% & $-$0&8\% & $-$0&8\% & $-$0&0\% & $-$0&9\% & $-$1&2\% & $-$0&5\% & $+$0&3\% \\
  production &  (1&3\%) &  (1&5\%) &  (1&6\%) &  (2&2\%) &  (1&7\%) &  (2&4\%) &  (2&5\%) &  (3&3\%) \\
\multicolumn{9}{c|}{ } & \multicolumn{8}{c}{ } \\
%BEMC ener-   & $-$7&2\% & $-$2&5\% & $-$0&6\% & $+$0&0\% & $-$6&0\% &  $-$1&0\% & $-$0&2\% & $+$0&0\% \\
BEMC ener-   & $-$2&1\% & $-$1&7\% & $-$0&8\% & $+$0&2\% & $-$0&4\% &  $-$1&1\% & $+$2&7\% & $+$2&4\% \\
gy scale     &  (1&2\%) &  (1&5\%) &  (1&5\%) &  (1&7\%) &  (1&4\%) &   (2&0\%) &  (2&2\%) &  (2&8\%) \\
\multicolumn{9}{c|}{ } & \multicolumn{8}{c}{ } \\
track    & $-$4&0\% & $-$3&7\% & $-$3&9\% & $-$3&1\% & $-$3&0\% & $-$3&5\% & $-$4&3\% & $-$4&5\% \\
selection&  (1&2\%) &  (1&5\%) &  (1&6\%) &  (1&5\%) &  (2&1\%) &  (1&9\%) &  (2&0\%) &  (2&4\%) \\
\multicolumn{9}{c|}{ } & \multicolumn{8}{c}{ } \\
unfolding    & $+$0&1\% & $-$0&8\% & $+$0&9\% & $+$1&8\% & $+$0&6\% & $-$1&2\% & $+$1&4\% & $+$1&1\% \\
  method     &  (1&2\%) &  (1&4\%) &  (1&7\%) &  (1&8\%) &  (1&7\%) &  (1&9\%) &  (2&4\%) &  (2&7\%) \\
\multicolumn{9}{c|}{ } & \multicolumn{8}{c}{ } \\
LEPTO 6.1    & $-$4&9\% & $-$4&5\% & $-$4&6\% & $-$3&8\% & $-$0&7\% & $-$2&3\% & $-$2&5\% & $-$2&1\% \\
             &  (1&6\%) &  (0&9\%) &  (1&2\%) &  (1&2\%) &  (1&1\%) &  (1&0\%) &  (1&2\%) &  (1&2\%) \\
\multicolumn{9}{c|}{ } & \multicolumn{8}{c}{ } \\
TOTAL        &    6&8\% &    6&6\% &    6&6\% &    5&9\% &    3&7\% &    5&3\% &    6&4\% &    6&5\% \\
\\
\end{tabular}}
\caption{Summary of systematic effects. The change in average multiplicity
is given for the full current hemisphere and for a restricted pseudorapidity
interval. For comparison, the  statistical error on $\aver{n}$ for the
given data sample is given in brackets.}
\label{syseffects}
\end{center}
\end{table}

 \begin{table}[p]
 \begin{center}
 {\scriptsize
 \begin{tabular}{ r 
 r@{.}l @{ $\pm$ } r@{.}l @{ $\pm$ } r@{.}l 
 r@{.}l @{ $\pm$ } r@{.}l @{ $\pm$ } r@{.}l 
 r@{.}l @{ $\pm$ } r@{.}l @{ $\pm$ } r@{.}l 
 r@{.}l @{ $\pm$ } r@{.}l @{ $\pm$ } r@{.}l }
 \multicolumn{25}{c}{\small 80~GeV $< W <$ 115~GeV} \\ \\
 $n$ & \multicolumn{24}{c}{$P_n$ (\%)} \\
 & \multicolumn{6}{c}{$1 < \eta^\ast < 2$     }
 & \multicolumn{6}{c}{$1 < \eta^\ast < 3$     }
 & \multicolumn{6}{c}{$1 < \eta^\ast < 4$     }
 & \multicolumn{6}{c}{$1 < \eta^\ast < 5$     }
 \\ 
 \hline \\
 0 & 15&79 &  0&52 &  2&18 &  3&26 &  0&22 &  1&58 &  0&78 &  0&09 &  1&66 &  0&40$^{\dag}$ &  0&\multicolumn{3}{l}{$\!\!\!\!08$} \\
 1 & 22&55 &  0&50 &  1&92 &  8&01 &  0&27 &  1&29 &  2&73 &  0&18 &  0&94 &  1&64 &  0&16 &  0&39 \\
 2 & 20&62 &  0&46 &  1&24 & 12&84 &  0&35 &  1&30 &  5&73 &  0&28 &  1&45 &  3&82 &  0&22 &  1&27 \\
 3 & 15&96 &  0&43 &  2&29 & 14&03 &  0&35 &  1&21 &  9&55 &  0&29 &  1&46 &  7&59 &  0&28 &  1&06 \\
 4 & 10&21 &  0&37 &  1&06 & 13&38 &  0&32 &  2&22 & 11&11 &  0&32 &  1&21 & 10&08 &  0&34 &  1&18 \\
 5 &  6&07 &  0&30 &  1&54 & 11&73 &  0&35 &  0&46 & 12&76 &  0&35 &  2&22 & 12&30 &  0&34 &  0&89 \\
 6 &  3&85 &  0&26 &  0&48 &  9&84 &  0&34 &  0&71 & 12&19 &  0&32 &  1&28 & 12&91 &  0&35 &  1&29 \\
 7 &  2&24 &  0&16 &  0&62 &  7&94 &  0&33 &  0&52 & 11&17 &  0&37 &  0&96 & 12&27 &  0&35 &  1&86 \\
 8 &  1&15 &  0&12 &  0&42 &  5&91 &  0&32 &  1&20 &  9&22 &  0&31 &  0&48 & 10&65 &  0&33 &  0&86 \\
 9 &  0&68 &  0&10 &  0&31 &  4&00 &  0&21 &  0&82 &  7&72 &  0&33 &  1&37 &  8&69 &  0&30 &  1&08 \\
10 &  0&49 &  0&12 &  0&24 &  3&21 &  0&19 &  0&56 &  5&69 &  0&24 &  1&31 &  6&38 &  0&26 &  0&99 \\
11 &  0&19 &  0&06 &  0&26 &  2&22 &  0&18 &  1&06 &  3&95 &  0&22 &  1&00 &  4&79 &  0&25 &  1&23 \\
12 &  0&07 &  0&04 &  0&13 &  1&49 &  0&14 &  0&66 &  2&76 &  0&23 &  0&96 &  3&20 &  0&24 &  0&97 \\
13 &  0&05 &  0&03 &  0&06 &  0&90 &  0&15 &  0&53 &  1&79 &  0&18 &  0&80 &  2&05 &  0&19 &  0&69 \\
14 & \multicolumn{6}{c}{ } &  0&54 &  0&11 &  0&53 &  1&22 &  0&15 &  0&62 &  1&38 &  0&16 &  0&74 \\
15 & \multicolumn{6}{c}{ } &  0&36 &  0&08 &  0&57 &  0&67 &  0&12 &  0&68 &  0&86 &  0&14 &  0&66 \\
16 & \multicolumn{6}{c}{ } &  0&14 &  0&07 &  0&19 &  0&40 &  0&11 &  0&43 &  0&40 &  0&08 &  0&24 \\
17 & \multicolumn{6}{c}{ } &  0&08 &  0&05 &  0&10 &  0&24 &  0&11 &  0&33 &  0&27 &  0&13 &  0&37 \\
18 & \multicolumn{6}{c}{ } & \multicolumn{6}{c}{ } &  0&13 &  0&08 &  0&19 &  0&10 &  0&05 &  0&11 \\
\\ 
\\
 \multicolumn{25}{c}{\small 115~GeV $< W <$ 150~GeV} \\ \\
 $n$ & \multicolumn{24}{c}{$P_n$ (\%)} \\
 & \multicolumn{6}{c}{$1 < \eta^\ast < 2$     }
 & \multicolumn{6}{c}{$1 < \eta^\ast < 3$     }
 & \multicolumn{6}{c}{$1 < \eta^\ast < 4$     }
 & \multicolumn{6}{c}{$1 < \eta^\ast < 5$     }
  \\ 
 \hline \\
 0 & 16&80 &  0&62 &  0&55 &  4&52 &  0&35 &  1&19 &  1&09 &  0&18 &  0&77 &  0&35$^{\dag}$ &  0&\multicolumn{3}{l}{$\!\!\!\!08$} \\
 1 & 21&55 &  0&54 &  1&05 &  8&00 &  0&35 &  1&21 &  2&57 &  0&19 &  1&12 &  1&17 &  0&13 &  0&62 \\
 2 & 19&69 &  0&50 &  0&53 & 12&15 &  0&36 &  0&70 &  4&72 &  0&24 &  0&70 &  2&58 &  0&14 &  1&26 \\
 3 & 15&19 &  0&48 &  0&77 & 12&96 &  0&36 &  0&91 &  8&09 &  0&29 &  0&74 &  5&63 &  0&23 &  0&69 \\
 4 & 10&54 &  0&36 &  0&31 & 12&74 &  0&38 &  0&74 &  9&96 &  0&33 &  0&95 &  7&97 &  0&24 &  1&03 \\
 5 &  6&76 &  0&34 &  0&75 & 11&09 &  0&32 &  1&63 & 11&23 &  0&36 &  0&95 & 10&36 &  0&31 &  0&82 \\
 6 &  4&17 &  0&27 &  0&50 &  9&35 &  0&31 &  0&70 & 11&27 &  0&33 &  1&37 & 11&44 &  0&37 &  0&56 \\
 7 &  2&55 &  0&20 &  0&76 &  7&31 &  0&27 &  2&06 & 10&32 &  0&31 &  0&25 & 11&96 &  0&35 &  1&18 \\
 8 &  1&29 &  0&14 &  0&48 &  6&01 &  0&29 &  0&93 &  9&21 &  0&34 &  0&52 & 10&41 &  0&32 &  0&66 \\
 9 &  0&64 &  0&10 &  0&62 &  4&53 &  0&24 &  1&38 &  8&03 &  0&30 &  0&29 &  9&65 &  0&36 &  0&74 \\
10 &  0&31 &  0&07 &  0&14 &  3&76 &  0&25 &  0&53 &  6&51 &  0&29 &  0&87 &  7&88 &  0&28 &  0&49 \\
11 &  0&20 &  0&04 &  0&07 &  2&57 &  0&23 &  0&86 &  5&32 &  0&26 &  0&88 &  6&15 &  0&23 &  0&80 \\
12 &  0&12 &  0&03 &  0&15 &  1&90 &  0&18 &  0&77 &  3&84 &  0&23 &  0&88 &  4&96 &  0&27 &  1&35 \\
13 &  0&07 &  0&03 &  0&07 &  1&05 &  0&16 &  0&29 &  2&96 &  0&23 &  0&53 &  3&51 &  0&25 &  0&50 \\
14 &  0&03 &  0&02 &  0&04 &  0&77 &  0&12 &  0&54 &  1&81 &  0&18 &  0&65 &  2&15 &  0&21 &  0&45 \\
15 & \multicolumn{6}{c}{ } &  0&56 &  0&10 &  0&48 &  1&25 &  0&16 &  0&33 &  1&47 &  0&17 &  0&43 \\
16 & \multicolumn{6}{c}{ } &  0&31 &  0&08 &  0&24 &  0&76 &  0&12 &  0&62 &  0&93 &  0&14 &  0&92 \\
17 & \multicolumn{6}{c}{ } &  0&18 &  0&06 &  0&28 &  0&46 &  0&10 &  0&29 &  0&72 &  0&17 &  0&62 \\
18 & \multicolumn{6}{c}{ } &  0&12 &  0&06 &  0&12 &  0&25 &  0&09 &  0&20 &  0&25 &  0&08 &  0&34 \\
19 & \multicolumn{6}{c}{ } & \multicolumn{6}{c}{ } &  0&14 &  0&06 &  0&23 &  0&17 &  0&07 &  0&24 \\
20 & \multicolumn{6}{c}{ } & \multicolumn{6}{c}{ } &  0&08 &  0&05 &  0&11 &  0&12 &  0&08 &  0&13 \\
21 & \multicolumn{6}{c}{ } & \multicolumn{6}{c}{ } & \multicolumn{6}{c}{ } &  0&03 &  0&02 &  0&03 \\
 \end{tabular}}
 \end{center}
 \label{tab:pnvsn}
 \caption{The fully corrected multiplicity distribution $P_n$ (\%).  The 
first error  is the statistical error, the  second the systematic 
uncertainty of the result. \hfil\break
$^{\dag}$ The value of $P_0$ in the domain $1 < \eta^{\ast} < 5$ is not measured 
but taken from  the reweighted DJANGO~6.0 Monte Carlo generator.}
 \end{table}

 \begin{table}[p]
 \begin{center}
 {\scriptsize
 \begin{tabular}{ r 
 r@{.}l @{ $\pm$ } r@{.}l @{ $\pm$ } r@{.}l 
 r@{.}l @{ $\pm$ } r@{.}l @{ $\pm$ } r@{.}l 
 r@{.}l @{ $\pm$ } r@{.}l @{ $\pm$ } r@{.}l 
 r@{.}l @{ $\pm$ } r@{.}l @{ $\pm$ } r@{.}l }
 \multicolumn{25}{c}{\small 150~GeV $< W <$ 185~GeV} \\ \\
 $n$ & \multicolumn{24}{c}{$P_n$ (\%)} \\
 & \multicolumn{6}{c}{$1 < \eta^\ast < 2$     }
 & \multicolumn{6}{c}{$1 < \eta^\ast < 3$     }
 & \multicolumn{6}{c}{$1 < \eta^\ast < 4$     }
 & \multicolumn{6}{c}{$1 < \eta^\ast < 5$     }
  \\ 
 \hline  \\
 0 & 16&01 &  0&58 &  5&43 &  4&24 &  0&36 &  3&19 &  1&13 &  0&15 &  1&51 &  0&28$^{\dag}$ &  0&\multicolumn{3}{l}{$\!\!\!\!08$} \\
 1 & 19&56 &  0&63 &  1&10 &  8&32 &  0&41 &  1&39 &  2&56 &  0&20 &  1&55 &  1&09 &  0&13 &  1&03 \\
 2 & 19&73 &  0&54 &  1&58 & 10&99 &  0&40 &  0&67 &  4&69 &  0&24 &  2&38 &  2&21 &  0&14 &  0&71 \\
 3 & 16&01 &  0&60 &  1&37 & 11&75 &  0&37 &  1&30 &  6&81 &  0&30 &  1&00 &  4&18 &  0&20 &  1&28 \\
 4 & 11&19 &  0&50 &  1&20 & 11&71 &  0&41 &  1&27 &  8&77 &  0&35 &  1&02 &  6&61 &  0&23 &  2&08 \\
 5 &  7&21 &  0&34 &  0&70 & 11&37 &  0&41 &  1&49 &  9&87 &  0&38 &  1&56 &  8&24 &  0&31 &  1&36 \\
 6 &  4&11 &  0&30 &  0&81 &  9&39 &  0&39 &  1&08 & 10&47 &  0&34 &  0&32 &  9&77 &  0&34 &  1&20 \\
 7 &  2&80 &  0&25 &  0&93 &  7&98 &  0&35 &  0&57 &  9&68 &  0&33 &  0&94 & 11&04 &  0&33 &  0&89 \\
 8 &  1&40 &  0&17 &  0&81 &  6&27 &  0&38 &  0&52 &  9&10 &  0&34 &  0&57 & 10&87 &  0&34 &  0&66 \\
 9 &  0&91 &  0&15 &  0&48 &  5&36 &  0&35 &  1&07 &  8&10 &  0&34 &  1&05 &  9&89 &  0&35 &  1&13 \\
10 &  0&33 &  0&08 &  0&23 &  3&64 &  0&25 &  1&03 &  6&90 &  0&31 &  0&81 &  8&18 &  0&35 &  0&44 \\
11 &  0&31 &  0&07 &  0&18 &  2&87 &  0&22 &  0&28 &  5&76 &  0&30 &  0&72 &  7&07 &  0&31 &  0&56 \\
12 &  0&10 &  0&04 &  0&10 &  1&82 &  0&20 &  0&73 &  4&92 &  0&28 &  1&17 &  5&76 &  0&34 &  1&15 \\
13 &  0&13 &  0&07 &  0&18 &  1&24 &  0&15 &  0&54 &  3&59 &  0&25 &  1&32 &  4&88 &  0&31 &  1&50 \\
14 &  0&08 &  0&06 &  0&14 &  1&08 &  0&16 &  0&89 &  2&71 &  0&24 &  0&97 &  3&42 &  0&26 &  0&89 \\
15 & \multicolumn{6}{c}{ } &  0&78 &  0&15 &  0&61 &  1&59 &  0&16 &  0&72 &  2&43 &  0&25 &  1&00 \\
16 & \multicolumn{6}{c}{ } &  0&43 &  0&09 &  0&40 &  1&11 &  0&12 &  0&93 &  1&31 &  0&17 &  0&78 \\
17 & \multicolumn{6}{c}{ } &  0&20 &  0&06 &  0&13 &  0&86 &  0&13 &  0&46 &  1&07 &  0&19 &  0&64 \\
18 & \multicolumn{6}{c}{ } &  0&18 &  0&07 &  0&13 &  0&49 &  0&13 &  0&32 &  0&55 &  0&14 &  0&51 \\
19 & \multicolumn{6}{c}{ } &  0&13 &  0&07 &  0&24 &  0&31 &  0&11 &  0&63 &  0&43 &  0&13 &  0&50 \\
20 & \multicolumn{6}{c}{ } &  0&06 &  0&05 &  0&10 &  0&09 &  0&05 &  0&14 &  0&12 &  0&05 &  0&51 \\
21 & \multicolumn{6}{c}{ } & \multicolumn{6}{c}{ } &  0&17 &  0&09 &  0&30 &  0&18 &  0&08 &  0&32 \\
22 & \multicolumn{6}{c}{ } & \multicolumn{6}{c}{ } &  0&13 &  0&09 &  0&26 &  0&16 &  0&11 &  0&31 \\
\\
\\
 \multicolumn{25}{c}{\small 185~GeV $< W <$ 220~GeV} \\ \\
 $n$ & \multicolumn{24}{c}{$P_n$ (\%)} \\
 & \multicolumn{6}{c}{$1 < \eta^\ast < 2$     }
 & \multicolumn{6}{c}{$1 < \eta^\ast < 3$     }
 & \multicolumn{6}{c}{$1 < \eta^\ast < 4$     }
 & \multicolumn{6}{c}{$1 < \eta^\ast < 5$     }
  \\ 
 \hline \\
 0 & 15&76 &  0&68 &  3&11 &  4&24 &  0&42 &  3&18 &  1&07 &  0&24 &  2&96 &  0&27$^{\dag}$ &  0&\multicolumn{3}{l}{$\!\!\!\!09$} \\
 1 & 19&66 &  0&76 &  0&78 &  8&00 &  0&45 &  0&48 &  2&81 &  0&30 &  1&27 &  0&96 &  0&13 &  1&96 \\
 2 & 19&56 &  0&62 &  1&75 & 10&74 &  0&57 &  1&52 &  5&22 &  0&32 &  0&73 &  2&37 &  0&23 &  0&68 \\
 3 & 15&76 &  0&61 &  2&03 & 11&51 &  0&47 &  0&69 &  6&70 &  0&33 &  0&50 &  3&73 &  0&24 &  1&12 \\
 4 & 11&08 &  0&58 &  1&85 & 11&76 &  0&48 &  2&08 &  7&98 &  0&37 &  0&87 &  5&47 &  0&27 &  0&99 \\
 5 &  6&95 &  0&45 &  0&86 & 11&65 &  0&53 &  1&27 &  9&18 &  0&34 &  0&58 &  7&50 &  0&30 &  0&83 \\
 6 &  4&63 &  0&38 &  0&97 &  9&74 &  0&47 &  1&62 &  9&07 &  0&40 &  0&87 &  8&49 &  0&35 &  1&24 \\
 7 &  3&20 &  0&32 &  1&14 &  7&96 &  0&43 &  0&91 & 10&18 &  0&50 &  0&63 & 10&08 &  0&39 &  1&61 \\
 8 &  1&36 &  0&16 &  0&74 &  6&83 &  0&44 &  1&06 &  9&10 &  0&41 &  0&68 & 10&54 &  0&46 &  1&18 \\
 9 &  0&95 &  0&15 &  0&79 &  5&12 &  0&34 &  1&16 &  7&95 &  0&34 &  1&57 & 10&40 &  0&43 &  1&71 \\
10 &  0&47 &  0&12 &  0&31 &  3&70 &  0&30 &  1&24 &  7&50 &  0&34 &  1&83 &  9&20 &  0&44 &  1&00 \\
11 &  0&33 &  0&10 &  0&50 &  2&52 &  0&24 &  0&88 &  6&10 &  0&38 &  0&78 &  8&03 &  0&40 &  1&12 \\
12 &  0&11 &  0&04 &  0&09 &  1&79 &  0&19 &  0&76 &  4&98 &  0&33 &  1&30 &  6&48 &  0&30 &  0&99 \\
13 &  0&05 &  0&04 &  0&10 &  1&39 &  0&17 &  0&92 &  3&46 &  0&26 &  0&77 &  4&78 &  0&29 &  0&95 \\
14 &  0&03 &  0&02 &  0&05 &  1&11 &  0&16 &  1&23 &  2&60 &  0&22 &  0&75 &  3&81 &  0&27 &  1&22 \\
15 & \multicolumn{6}{c}{ } &  0&66 &  0&13 &  0&33 &  2&03 &  0&21 &  0&58 &  2&32 &  0&21 &  0&78 \\
16 & \multicolumn{6}{c}{ } &  0&44 &  0&10 &  0&27 &  1&29 &  0&16 &  0&80 &  2&13 &  0&22 &  1&08 \\
17 & \multicolumn{6}{c}{ } &  0&29 &  0&10 &  0&19 &  1&00 &  0&16 &  0&65 &  1&14 &  0&14 &  0&67 \\
18 & \multicolumn{6}{c}{ } &  0&14 &  0&06 &  0&19 &  0&71 &  0&15 &  0&22 &  0&83 &  0&12 &  0&44 \\
19 & \multicolumn{6}{c}{ } &  0&10 &  0&05 &  0&14 &  0&39 &  0&09 &  0&35 &  0&64 &  0&14 &  0&29 \\
20 & \multicolumn{6}{c}{ } &  0&06 &  0&04 &  0&29 &  0&21 &  0&07 &  0&18 &  0&30 &  0&07 &  0&39 \\
21 & \multicolumn{6}{c}{ } &  0&05 &  0&04 &  0&10 &  0&17 &  0&10 &  0&26 &  0&13 &  0&07 &  0&18 \\
22 & \multicolumn{6}{c}{ } & \multicolumn{6}{c}{ } &  0&11 &  0&04 &  0&13 &  0&19 &  0&08 &  0&26 \\
23 & \multicolumn{6}{c}{ } & \multicolumn{6}{c}{ } & \multicolumn{6}{c}{ } &  0&03 &  0&02 &  0&05 \\
 \end{tabular}}
 \end{center}
 \addtocounter{table}{-1}
 \caption{continued}
 \end{table}

\begin{table}[p]
\begin{center}
{\scriptsize
\begin{tabular}
{ 
c 
r@{.}l @{ $\pm$ } r@{.}l @{ $\pm$ } r@{.}l
r@{.}l @{ $\pm$ } r@{.}l @{ $\pm$ } r@{.}l
r@{.}l @{ $\pm$ } r@{.}l @{ $\pm$ } r@{.}l
r@{.}l @{ $\pm$ } r@{.}l @{ $\pm$ } r@{.}l }
\multicolumn{25}{c}{\small $\eta^\ast>0$}
\\ \\
$W$ (GeV) &
\multicolumn{6}{c}{$80 \div 115$} &
\multicolumn{6}{c}{$115 \div 150$} &
\multicolumn{6}{c}{$150 \div 185$} &
\multicolumn{6}{c}{$185 \div 220$} \\
 \hline \\
$\anch$ & 8&98 &  0&07 &  0&61 & 10&00 &  0&07 &  0&66 & 10&88 &  0&09 &  0&72 & 11&35 &  0&09 &  0&67 \\
$D_2$ & 3&89 &  0&07 &  0&23 &  4&16 &  0&06 &  0&26 &  4&49 &  0&08 &  0&36 &  4&56 &  0&08 &  0&30 \\
$D_3$ & 3&43 &  0&16 &  0&22 &  3&41 &  0&15 &  0&21 &  3&61 &  0&26 &  0&55 &  3&67 &  0&25 &  0&45 \\
$D_4$ & 5&39 &  0&15 &  0&38 &  5&58 &  0&13 &  0&23 &  6&05 &  0&21 &  0&71 &  6&26 &  0&19 &  0&49 \\
$C_2$ & 1&18 &  0&00 &  0&04 &  1&17 &  0&00 &  0&01 &  1&17 &  0&00 &  0&02 &  1&16 &  0&00 &  0&02 \\
$C_3$ & 1&62 &  0&02 &  0&14 &  1&56 &  0&01 &  0&06 &  1&54 &  0&02 &  0&08 &  1&51 &  0&02 &  0&07 \\
$C_4$ & 2&48 &  0&07 &  0&40 &  2&29 &  0&05 &  0&17 &  2&26 &  0&07 &  0&21 &  2&19 &  0&06 &  0&19 \\
$R_2$ & 1&07 &  0&00 &  0&03 &  1&07 &  0&00 &  0&01 &  1&07 &  0&00 &  0&02 &  1&07 &  0&00 &  0&02 \\
$R_3$ & 1&24 &  0&02 &  0&10 &  1&22 &  0&01 &  0&04 &  1&24 &  0&02 &  0&07 &  1&22 &  0&02 &  0&06 \\
\\
\\
\multicolumn{25}{c}{{\small $1 < \eta^\ast < 2$}} 
\\ \\
$W$ (GeV) &
\multicolumn{6}{c}{$80 \div 115$} &
\multicolumn{6}{c}{$115 \div 150$} &
\multicolumn{6}{c}{$150 \div 185$} &
\multicolumn{6}{c}{$185 \div 220$} \\
 \hline \\
$\anch$ & 2&46 &  0&03 &  0&10 &  2&50 &  0&03 &  0&12 &  2&63 &  0&04 &  0&18 &  2&66 &  0&04 &  0&18 \\
$D_2$ & 2&07 &  0&05 &  0&08 &  2&13 &  0&03 &  0&10 &  2&22 &  0&05 &  0&12 &  2&22 &  0&03 &  0&14 \\
$D_3$ & 2&21 &  0&11 &  0&10 &  2&24 &  0&05 &  0&14 &  2&39 &  0&11 &  0&18 &  2&31 &  0&07 &  0&12 \\
$D_4$ & 3&08 &  0&16 &  0&15 &  3&15 &  0&07 &  0&17 &  3&39 &  0&16 &  0&26 &  3&24 &  0&09 &  0&12 \\
$C_2$ & 1&71 &  0&02 &  0&02 &  1&72 &  0&02 &  0&04 &  1&71 &  0&02 &  0&05 &  1&69 &  0&02 &  0&03 \\
$C_3$ & 3&87 &  0&16 &  0&18 &  3&90 &  0&11 &  0&24 &  3&89 &  0&17 &  0&18 &  3&75 &  0&12 &  0&20 \\
$C_4$ &10&67 &  0&93 &  1&23 & 10&77 &  0&60 &  1&17 & 11&03 &  1&09 &  0&63 & 10&02 &  0&65 &  1&22 \\
$R_2$ & 1&30 &  0&02 &  0&03 &  1&32 &  0&01 &  0&04 &  1&33 &  0&02 &  0&04 &  1&32 &  0&02 &  0&03 \\
$R_3$ & 2&11 &  0&14 &  0&15 &  2&15 &  0&08 &  0&19 &  2&22 &  0&14 &  0&14 &  2&11 &  0&09 &  0&13 \\
$K_3$ &  0&187 &  0&070 &  0&097 &   0&167 &  0&040 &  0&063 &   0&226 &  0&079 &  0&085 &  0&147 &  0&043 &
0&101 \\
\\
\\
\multicolumn{25}{c}{{\small $1 < \eta^\ast < 3$}} 
\\ \\
$W$ (GeV) &
\multicolumn{6}{c}{$80 \div 115$} & 
\multicolumn{6}{c}{$115 \div 150$} &
\multicolumn{6}{c}{$150 \div 185$} & 
\multicolumn{6}{c}{$185 \div 220$} \\
\hline \\
$\anch$ & 4&90 &  0&04 &  0&18 &  5&06 &  0&05 &  0&27 &  5&32 &  0&07 &  0&34 & 5&35 &  0&06 &  0&35 \\
$D_2$ & 3&10 &  0&04 &  0&13 &  3&37 &  0&04 &  0&21 &  3&57 &  0&06 &  0&26 & 3&58 &  0&07 &  0&30 \\
$D_3$ & 2&91 &  0&10 &  0&25 &  3&18 &  0&09 &  0&21 &  3&45 &  0&12 &  0&43 & 3&54 &  0&20 &  0&34 \\
$D_4$ & 4&27 &  0&11 &  0&28 &  4&62 &  0&10 &  0&27 &  5&01 &  0&13 &  0&45 & 5&18 &  0&28 &  0&45 \\
$C_2$ & 1&40 &  0&01 &  0&03 &  1&44 &  0&01 &  0&02 &  1&45 &  0&01 &  0&02 & 1&44 &  0&01 &  0&02 \\
$C_3$ & 2&41 &  0&04 &  0&16 &  2&57 &  0&05 &  0&10 &  2&62 &  0&07 &  0&14 & 2&62 &  0&09 &  0&09 \\
$C_4$ & 4&83 &  0&19 &  0&64 &  5&34 &  0&20 &  0&38 &  5&57 &  0&28 &  0&68 & 5&71 &  0&47 &  0&34 \\
$R_2$ & 1&19 &  0&01 &  0&03 &  1&24 &  0&01 &  0&02 &  1&26 &  0&01 &  0&02 & 1&26 &  0&01 &  0&03 \\
$R_3$ & 1&64 &  0&04 &  0&13 &  1&80 &  0&04 &  0&09 &  1&87 &  0&06 &  0&13 & 1&88 &  0&08 &  0&11 \\
$K_3$ &  0&048 &  0&015 &  0&044 &   0&063 &  0&016 &  0&033 &   0&088 &  0&022 &  0&077 &  0&108 &  0&041 &
0&036 \\
\\
\\
\multicolumn{25}{c}{{\small $1 < \eta^\ast < 4$}}
\\ \\
$W$ (GeV) &
\multicolumn{6}{c}{$80 \div 115$} &
\multicolumn{6}{c}{$115 \div 150$} &
\multicolumn{6}{c}{$150 \div 185$} &
\multicolumn{6}{c}{$185 \div 220$} \\
\hline \\
$\anch$ & 6&45 &  0&06 &  0&33 &  7&00 &  0&05 &  0&35 &  7&51 &  0&06 &  0&51 & 7&66 &  0&07 &  0&47 \\
$D_2$ & 3&23 &  0&09 &  0&14 &  3&57 &  0&04 &  0&23 &  3&93 &  0&06 &  0&25 & 4&06 &  0&06 &  0&20 \\
$D_3$ & 2&71 &  0&28 &  0&21 &  2&89 &  0&11 &  0&23 &  3&28 &  0&14 &  0&50 & 3&32 &  0&16 &  0&28 \\
$D_4$ & 4&34 &  0&27 &  0&24 &  4&72 &  0&10 &  0&20 &  5&29 &  0&14 &  0&44 & 5&41 &  0&14 &  0&22 \\
$C_2$ & 1&25 &  0&01 &  0&03 &  1&26 &  0&00 &  0&02 &  1&27 &  0&00 &  0&02 & 1&28 &  0&00 &  0&02 \\
$C_3$ & 1&82 &  0&05 &  0&13 &  1&85 &  0&02 &  0&07 &  1&90 &  0&03 &  0&08 & 1&92 &  0&03 &  0&08 \\
$C_4$ & 3&01 &  0&20 &  0&38 &  3&04 &  0&08 &  0&22 &  3&22 &  0&11 &  0&27 & 3&26 &  0&12 &  0&26 \\
$R_2$ & 1&09 &  0&01 &  0&03 &  1&11 &  0&00 &  0&02 &  1&14 &  0&00 &  0&02 & 1&15 &  0&00 &  0&02 \\
$R_3$ & 1&29 &  0&05 &  0&09 &  1&35 &  0&02 &  0&06 &  1&43 &  0&03 &  0&07 & 1&45 &  0&03 &  0&06 \\
\end{tabular}}
\end{center}
\caption{Moments and cumulants of the unfolded multiplicity distribution for different ranges in 
pseudorapidity $\eta^\ast$ and intervals in $W$.}
\label{tab:moments}
\end{table}

\begin{table}[p]
\begin{center}
{\scriptsize
\begin{tabular}
{ 
c 
r@{.}l @{ $\pm$ } r@{.}l @{ $\pm$ } r@{.}l
r@{.}l @{ $\pm$ } r@{.}l @{ $\pm$ } r@{.}l
r@{.}l @{ $\pm$ } r@{.}l @{ $\pm$ } r@{.}l
r@{.}l @{ $\pm$ } r@{.}l @{ $\pm$ } r@{.}l }
\multicolumn{25}{c}{{\small $1 < \eta^\ast < 5$}}
\\ \\
$W$ (GeV) &
\multicolumn{6}{c}{$80 \div 115$} &
\multicolumn{6}{c}{$115 \div 150$} &
\multicolumn{6}{c}{$150 \div 185$} &
\multicolumn{6}{c}{$185 \div 220$} \\
\hline \\
$\anch$ & 6&90 &  0&05 &  0&34 &  7&72 &  0&05 &  0&42 &  8&45 &  0&06 &  0&58 & 8&81 &  0&08 &  0&54 \\
$D_2$ & 3&15 &  0&05 &  0&21 &  3&45 &  0&04 &  0&25 &  3&77 &  0&06 &  0&25 & 3&86 &  0&05 &  0&22 \\
$D_3$ & 2&56 &  0&12 &  0&33 &  2&74 &  0&13 &  0&19 &  3&02 &  0&18 &  0&54 & 2&95 &  0&19 &  0&41 \\
$D_4$ & 4&23 &  0&11 &  0&37 &  4&59 &  0&10 &  0&29 &  5&08 &  0&15 &  0&45 & 5&17 &  0&14 &  0&29 \\
$C_2$ & 1&20 &  0&00 &  0&03 &  1&19 &  0&00 &  0&02 &  1&19 &  0&00 &  0&02 & 1&19 &  0&00 &  0&02 \\
$C_3$ & 1&67 &  0&02 &  0&12 &  1&64 &  0&02 &  0&06 &  1&64 &  0&02 &  0&08 & 1&61 &  0&02 &  0&08 \\
$C_4$ & 2&60 &  0&07 &  0&35 &  2&50 &  0&06 &  0&16 &  2&50 &  0&08 &  0&24 & 2&42 &  0&07 &  0&21 \\
$R_2$ & 1&06 &  0&00 &  0&02 &  1&07 &  0&00 &  0&02 &  1&08 &  0&00 &  0&02 & 1&07 &  0&00 &  0&02 \\
$R_3$ & 1&19 &  0&02 &  0&09 &  1&21 &  0&01 &  0&05 &  1&24 &  0&02 &  0&07 & 1&23 &  0&02 &  0&06 \\
\\
\\
\multicolumn{25}{c}{{\small $2 < \eta^\ast < 3$}}
\\ \\
$W$ (GeV) &
\multicolumn{6}{c}{$80 \div 115$} &
\multicolumn{6}{c}{$115 \div 150$} &
\multicolumn{6}{c}{$150 \div 185$} &
\multicolumn{6}{c}{$185 \div 220$} \\
 \hline \\
$\anch$ & 2&46 &  0&03 &  0&17 &  2&61 &  0&03 &  0&20 &  2&73 &  0&04 &  0&27 & 2&71 &  0&04 &  0&20 \\
$D_2$ & 1&92 &  0&03 &  0&19 &  2&11 &  0&03 &  0&16 &  2&24 &  0&04 &  0&20 & 2&22 &  0&03 &  0&18 \\
$D_3$ & 1&86 &  0&05 &  0&10 &  2&07 &  0&06 &  0&14 &  2&23 &  0&08 &  0&13 & 2&28 &  0&07 &  0&22 \\
$D_4$ & 2&67 &  0&06 &  0&12 &  2&94 &  0&08 &  0&21 &  3&14 &  0&09 &  0&18 & 3&20 &  0&09 &  0&25 \\
$C_2$ & 1&61 &  0&01 &  0&03 &  1&65 &  0&01 &  0&03 &  1&67 &  0&01 &  0&05 & 1&67 &  0&02 &  0&06 \\
$C_3$ & 3&26 &  0&08 &  0&16 &  3&47 &  0&10 &  0&15 &  3&57 &  0&10 &  0&25 & 3&60 &  0&11 &  0&30 \\
$C_4$ & 7&77 &  0&35 &  0&79 &  8&57 &  0&48 &  0&60 &  8&99 &  0&52 &  1&20 & 9&30 &  0&56 &  1&19 \\
$R_2$ & 1&20 &  0&01 &  0&02 &  1&27 &  0&01 &  0&04 &  1&30 &  0&01 &  0&05 & 1&30 &  0&02 &  0&06 \\
$R_3$ & 1&63 &  0&05 &  0&10 &  1&86 &  0&08 &  0&13 &  1&99 &  0&08 &  0&14 & 2&02 &  0&09 &  0&25 \\
$K_3$ &  0&018 &  0&021 &  0&041 &   0&043 &  0&030 &  0&035 &   0&072 &  0&040 &  0&062 &  0&121 &  0&038 &
0&076 \\
\\
\\
\multicolumn{25}{c}{{\small $3 < \eta^\ast < 4$}}
\\ \\
$W$ (GeV) &
\multicolumn{6}{c}{$80 \div 115$} &
\multicolumn{6}{c}{$115 \div 150$} &
\multicolumn{6}{c}{$150 \div 185$} &
\multicolumn{6}{c}{$185 \div 220$} \\
\hline \\
$\anch$ & 1&54 &  0&02 &  0&18 &  1&93 &  0&02 &  0&17 &  2&17 &  0&02 &  0&18 & 2&29 &  0&03 &  0&17 \\
$D_2$ & 1&32 &  0&01 &  0&14 &  1&48 &  0&02 &  0&14 &  1&59 &  0&02 &  0&09 & 1&67 &  0&03 &  0&10 \\
$D_3$ & 1&23 &  0&03 &  0&07 &  1&33 &  0&03 &  0&09 &  1&39 &  0&05 &  0&04 & 1&54 &  0&05 &  0&16 \\
$D_4$ & 1&79 &  0&03 &  0&11 &  2&01 &  0&04 &  0&10 &  2&14 &  0&06 &  0&06 & 2&31 &  0&06 &  0&17 \\
$C_2$ & 1&73 &  0&01 &  0&11 &  1&58 &  0&01 &  0&10 &  1&53 &  0&01 &  0&06 & 1&53 &  0&01 &  0&10 \\
$C_3$ & 3&70 &  0&08 &  0&55 &  3&08 &  0&07 &  0&47 &  2&87 &  0&07 &  0&26 & 2&91 &  0&08 &  0&46 \\
$C_4$ & 9&24 &  0&38 &  2&29 &  6&99 &  0&28 &  1&72 &  6&21 &  0&31 &  1&00 & 6&47 &  0&34 &  1&70 \\
$R_2$ & 1&08 &  0&01 &  0&04 &  1&07 &  0&01 &  0&07 &  1&07 &  0&01 &  0&03 & 1&09 &  0&01 &  0&08 \\
$R_3$ & 1&17 &  0&05 &  0&15 &  1&16 &  0&04 &  0&19 &  1&18 &  0&05 &  0&08 & 1&28 &  0&05 &  0&24 \\
$K_3$ &  0&079 &  0&019 &  0&062 &   0&045 &  0&014 &  0&040 &   0&054 &  0&018 &  0&033 &  0&011 &  0&017 &
0&068 \\
\\
\\
\multicolumn{25}{c}{{\small $4 < \eta^\ast < 5$}}
\\ \\
$W$ (GeV) &
\multicolumn{6}{c}{$80 \div 115$} &
\multicolumn{6}{c}{$115 \div 150$} &
\multicolumn{6}{c}{$150 \div 185$} &
\multicolumn{6}{c}{$185 \div 220$} \\
\hline \\
$\anch$ & 0&46 &  0&01 &  0&04 &  0&71 &  0&01 &  0&04 &  0&93 &  0&01 &  0&09 & 1&17 &  0&01 &  0&09 \\
$D_2$ & 0&71 &  0&01 &  0&02 &  0&88 &  0&01 &  0&08 &  0&99 &  0&01 &  0&06 & 1&09 &  0&01 &  0&19 \\
$D_3$ & 0&83 &  0&02 &  0&02 &  0&93 &  0&01 &  0&03 &  0&98 &  0&02 &  0&05 & 1&01 &  0&02 &  0&01 \\
$D_4$ & 1&09 &  0&03 &  0&04 &  1&26 &  0&02 &  0&03 &  1&37 &  0&03 &  0&07 & 1&47 &  0&02 &  0&03 \\
$C_2$ & 3&34 &  0&06 &  0&11 &  2&52 &  0&04 &  0&18 &  2&13 &  0&02 &  0&09 & 1&86 &  0&01 &  0&11 \\
$C_3$ &13&78 &  0&61 &  1&06 &  7&85 &  0&30 &  1&18 &  5&58 &  0&17 &  0&59 & 4&23 &  0&10 &  0&55 \\
$C_4$ &68&80 &  5&70 &  9&55 & 28&93 &  1&85 &  6&83 & 17&21 &  0&87 &  3&47 &11&24 &  0&46 &  2&25 \\
$R_2$ & 1&19 &  0&04 &  0&06 &  1&12 &  0&03 &  0&10 &  1&05 &  0&02 &  0&05 & 1&01 &  0&01 &  0&05 \\
$R_3$ & 1&41 &  0&24 &  0&29 &  1&16 &  0&09 &  0&23 &  1&02 &  0&07 &  0&19 & 0&92 &  0&04 &  0&09 \\
$K_3$ &  0&159 &  0&144 &  0&138 &   0&221 &  0&041 &  0&136 &   0&150 &  0&028 &  0&104 &  0&109 &  0&017 &
0&061 \\
\end{tabular}}
\end{center}
\addtocounter{table}{-1}
\caption{continued}
\end{table}

%% file: multpap.bbl
\begin{thebibliography}{10}

\bibitem{mul:exp:rev}
{{\em Multiparticle Dynamics\/} (Festschrift L. Van Hove), Eds.~A.~Giovannini
  and W.Kittel (World Scientific, Singapore 1990);\\ G.~Giacomelli, Int. J.
  Mod. Phys. {\rm A5} (1990) 223}.

\bibitem{schmitz:rev}
{For comprehensive reviews see N.~Schmitz, Int. J. Mod. Phys. {\rm A3} (1988)
  1997; {\em ibid.} {\rm A8} (1993) 1993}.

\bibitem{carr:shih}
{P.~Carruthers and C.C.~Shih, Int. J. Mod. Phys. {\rm A2} (1987) 1447}.

\bibitem{dremin:rev}
{I.M.~Dremin, Sov. Phys. Uspekhi {\rm37} (1994) 4077}.

\bibitem{ugoccioni}
{R.~Ugoccioni, A.~Giovannini and S.~Lupia, in {\MPVIETRI} p.384}.

\bibitem{heisenberg:fermi}
{W.~Heisenberg, Z.Phys. {\rm113} (1939) 61; Nature (London) {\rm 164} (1949)
  65;\\ E.~Fermi, Prog. Theor. Phys. {\rm5} (1950) 570}.

\bibitem{bialas:hayot}
{A.~Bia\l as and F.~Hayot, {\PRV{D33}{1986}{39}}}.

\bibitem{review:93b}
{E.A.~De Wolf, I.M.~Dremin and W.~Kittel, Phys. Rep. {\rm 270} (1996) 1}.

\bibitem{zeus:md}
{ZEUS Collaboration, M.~Derrick {\etal}, {\ZF{67}{1995}{93}}}.

\bibitem{h1:thompson}
{H1 Collaboration, S.~Aid {\etal}, {\NP{B445}{1995}{3}}}.

\bibitem{Mue71}
{A.H. Mueller, {\PRV{D4}{1971}{150}}}.

\bibitem{KNO}
{Z.Koba, H.B.~Nielsen and P.~Olesen, \NP{B40}{1972}{317}}.

\bibitem{feynman:69}
{R.P.~Feynman, {\PRL{23}{1969}{1415}}}.

\bibitem{golov}
{A.I.~Golokhvastov, {\SJNP{27}{1978}{430}}; {\em ibid.\/} {\rm 30} (1979) 128}.

\bibitem{ochs:md:90}
{For a review see W.~Ochs, in \MPHOLMECKE\ p.~434.}

\bibitem{wrob:logn}
{ S.~Carius and G.~Ingelman, {\PL{B252}{1990}{647}}; \\ G.~Wrochna, {\em How to
  fit the Lognormal\/}, preprint Univ. of Warsaw, IFD/8/1990 (1990);\\
  R.~Szwed, G.~Wrochna and A.K.~Wr\'oblewski, {\MPL{A6}{1991}{245}}; \\
  M.Gazdzicki {\etal}, {\MPL{A6}{1991}{981}}}.

\bibitem{szwed:wrochna:85}
{R.~Szwed and G.~Wrochna, {\ZF{29}{1985}{255}}}.

\bibitem{Carl87:2}
{UA5 Collaboration, G.J. Alner {\etal}, Phys.~Rep. {\rm154} (1987) 247}.

\bibitem{ua5:md:2:9}
{UA5 Collaboration, R.E.~Ansorge {\it et al.\/}, {\ZF{43}{1989}{357}}}.

\bibitem{scale:cascade}
{A.M.~Polyakov Sov. Phys.--JETP {\rm32} (1971) 296; {\rm33} (1971) 850; \\
  S.J.~Orfanidis and V.~Rittenberg, {\PRV{D10}{1974}{2892}};\\
  G.~Cohen-Tannoudji and W.~Ochs, {\ZF{39}{1988}{513}};\\ W.~Ochs,
  {\ZF{23}{1984}{131}}}.

\bibitem{vanhove:1}
{L.~Van Hove and A.~Giovannini, Acta. Phys. Pol. {\rm B19} (1988) 917;\\
  A.~Giovannini and L.~Van Hove, {\ZF{30}{1986}{391}}}.

\bibitem{Giovan79}
{A. Giovannini, {\NP{B161}{1979}{429}}}.

\bibitem{lphd}
{D.~Amati and G.~Veneziano, {\PL{B83}{1979}{87}};\\ G.~Marchesini, L.~Trentadue
  and G.~Veneziano, {\NP{B181}{1981}{335}};\\ Ya.I.~Azimov {\etal},
  {\ZF{27}{1985}{65}}; {\em ibid.\/} {\rm31} (1986) 213}.

\bibitem{malaza:webber}
{E.D.~Malaza and B.R.~Webber, {\NP{B267}{1986}{702}}; {\em ibid.}
  {\PL{B149}{1984}{510}}}.

\bibitem{dipole:model}
{G.~Gustafson, {\NP{B392}{1993}{251}};\\ G.~Gustafson and M.~Olsson,
  {\NP{B406}{1993}{293}}}.

\bibitem{doksh:93}
{Yu.L.~Dokshitzer, {\PL{B305}{1993}{295}}}.

\bibitem{delphi:md}
{DELPHI Collaboration, P.~Abreu {\etal}, {\ZF{50}{1991}{185}}; {\em ibid.} {\rm
  C52} (1991) 271; {\em ibid.} {\rm C56} (1992) 63}.

\bibitem{aleph:rap:mul}
{ALEPH Collaboration, D.~Buskulic {\etal}, {\ZF{69}{1995}{1}}}.

\bibitem{e665:mul}
{E665 Collaboration, M.R.~Adams {\etal}, {\ZF{61}{1994}{179}}}.

\bibitem{wrob:law}
{A.K.~Wr\'oblewski, {\APP{B4}{1973}{857}};\\ A.J.~Buras {\etal},
  {\PL{B47}{1973}{251}}}.

\bibitem{opal:mul}
{OPAL Collaboration, P.D.~Acton {\etal} {\ZF{53}{1992}{539}}}.

\bibitem{aleph:mul}
{ALEPH Collaboration, P.~Decamp {\etal}, {\PL{B273}{1991}{181}}}.

\bibitem{jones:92}
{WA21 Collaboration, C.T.~Jones {\etal}, {\ZF{54}{1992}{45}}}.

\bibitem{h1:detector}
{H1 Collaboration, I.~Abt {\etal}, DESY 93-103 (1993)}.

\bibitem{h1:calo}
{H1 Calorimeter Group, B.~Andrieu {\etal}, {\NIM{A336}{1993}{460}}}.

\bibitem{h1:rapgap}
{H1 Collaboration, T.~Ahmed {\etal}, {\NP{B249}{1994}{477}}}.

\bibitem{DJANGO}
{K.~Charchula, G.~Schuler and H.~Spiesberger, CERN-TH.7133/94.}

\bibitem{HERACLES}
{A.~Kwiatkowski, H.~Spiesberger and H.-J.~M\"ohring, Comp. Phys. Comm. {\rm69}
  (1992) 155}.

\bibitem{LEPTO}
{G.~Ingelman, Proceedings of the 1991 Workshop on Physics at HERA, DESY Vol.3
  (1992) 1366.}

\bibitem{GRV}
{M.~Gl\"uck, E.~Reya and A.~Vogt, {\ZF{67}{1995}{433}}}.

\bibitem{ARIADNE}
{L.~L\"onnblad, Comp. Phys. Comm. {\rm71} (1992) 15.}

\bibitem{JETSET}
{T.~Sj\"ostrand, Comp. Phys. Comm. {\rm39} (1986) 346;\\ T.~Sj\"ostrand and
  M.~Bengtsson, {\em ibid.\/} 43 (1987) 367.}

\bibitem{lundstring}
{B.~Andersson {\etal}, {\PR{97}{1983}{31}}}.

\bibitem{HERWIG}
{G.~Marchesini {\etal}, Comp. Phys. Comm. {\rm67} (1992) 465}.

\bibitem{GEANT}
{R.~Brun {\etal}, GEANT3, CERN DD/EE/84-1 (1987).}

\bibitem{tasso:mul}
{TASSO Collaboration, W.~Braunschweig {\etal} {\ZF{45}{1989}{193}}}.

\bibitem{amy:mul}
{AMY Collaboration, H.W.~Zheng {\etal}, {\PRV{D42}{1990}{737}}}.

\bibitem{delphi:mul}
{DELPHI Collaboration, P.~Abreu {\etal}, {\ZF{52}{1991}{271}};
  {\ZF{56}{1992}{63}}}.

\bibitem{h1:f2_96}
{H1 Collaboration, S.~Aid {\etal}, {\NP{B470}{1996}{3}}}.

\bibitem{zeus:kzero}
{ZEUS Collaboration, M.~Derrick {\etal}, {\ZF{68}{1995}{29}}}.

\bibitem{h1:kzero}
{H1 Collaboration, S.~Aid {\etal}, DESY 96-122 (1996) and hep-ex/9607010}.

\bibitem{hrs:mul}
{HRS Collaboration, M.~Derrick {\etal}, {\PRV{D34}{1986}{3304}}}.

\bibitem{delphi:jet:tune}
{DELPHI Collaboration, P.~Abreu {\etal}, {\em Tuning and test of fragmentation
  models based on identified particles and precision event shape data\/},
  contributed paper eps0548 to EPS-HEP95, Brussels 1995}.

\bibitem{delphi:md:130}
{DELPHI Collaboration, P.~Abreu {\etal}, {\PL{B372}{1996}{172}}}.

\bibitem{delphi:charm}
{A.~De Angelis, in Proc. Int. Europhysics Conf. on High Energy Physics,
  Brussels 1995, Eds. J.~Lemonne, C.~Vander Velde and F.~Verbeure (World
  Scientific Singapore, 1996) p.~63}.

\bibitem{jongej:89}
{WA25 Collaboration, B.~Jongejans {\etal}, {\NC{A101}{1989}{435}}}.

\bibitem{emc:mul}
{EMC Collaboration, M.~Arneodo {\etal}, {\ZF{35}{1987}{335}}}.

\bibitem{doksh:91}
{Yu.L.~Dokshitzer {\etal}, {\em Basics of Perturbative QCD\/} (Editions
  Fronti\`eres, Gif-sur-Yvette, 1991)}.

\bibitem{lupia:ochs}
{S.~Lupia and W.~Ochs, {\PL{B365}{1996}{334}}}.

\bibitem{webber:84}
{B.R.~Webber, {\PL{143B}{1984}{501}} and refs. therein}.

\bibitem{virchaux}
{M.~Virchaux and A.~Milszstajn, {\PL{B274}{1992}{221}}}.

\bibitem{emc:85np}
{EMC Collaboration, M.~Arneodo {\etal}, {\NP{B258}{1985}{249}}}.

\bibitem{ochs:vietri}
{W.~Ochs, in \MPVIETRI\ p.~243.}

\bibitem{kant:thesis}
{D.~Kant, University of London thesis, also published by the Rutherford
  Appleton Laboratory as RAL-TH-96-008}.

\bibitem{emc:ee}
{EMC Collaboration, M.~Arneodo {\etal}, {\ZF{35}{1987}{417}}}.

\bibitem{e665:spectra:91}
{E665 Collaboration, M.R.~Adams {\etal}, {\PL{B272}{1991}{163}}}.

\bibitem{h1:eflow:94}
{H1 Collaboration, I.~Abt {\etal}, {\ZF{63}{1994}{377}}}.

\bibitem{zeus:single:95}
{ZEUS Collaboration, M.~Derrick {\etal}, {\ZF{70}{1996}{1}}}.

\bibitem{opal:corr:xi}
{OPAL Collaboration, P.D.~Acton {\etal}, {\PL{B287}{1993}{401}}}.

\bibitem{noqvar}
{ P.H.~Garbincius {\etal}, {\PRL{32}{1974}{328}};\\ A.J.~Sadoff {\etal},
  {\PRL{32}{1974}{955}};\\ J.~Ballam {\etal}, {\PL{56B}{1975}{193}};\\
  B.~Gibbard {\etal}, {\PRV{D11}{1975}{2367}};\\ C.~del Papa {\etal},
  {\PRV{D13}{1976}{2934}};\\ C.K.~Chen {\etal}, {\NP{B133}{1978}{13}};\\
  M.~Derrick {\etal}, {\PL{91B}{1980}{470}};\\ P.~Allen {\etal},
  {\NP{B181}{1981}{385}};\\ EMC Collaboration, J.J.~Aubert {\etal},
  {\PL{114B}{1982}{373}};\\ V.V.~Amossov {\etal}, {\NP{B203}{1982}{1}};\\
  B.~Barlag {\etal}, {\ZF{11}{1982}{283}}, {\rm C14} (1982) 281, erratum;\\
  H.~Gr\"assler {\etal}, {\NP{B233}{1983}{269}};\\ EMC Collaboration,
  M.~Arneodo {\etal}, {\PL{165B}{1985}{222}}}.

\bibitem{bj:kogut}
{J.D.~Bjorken and J.~Kogut, {\PRV{D8}{1973}{1341}}}.

\bibitem{cahn:cleymans}
{R.N.~Cahn, J.W.~Cleymans and E.W.~Colglazier,{\PL{34B}{1973}{323}}}.

\bibitem{emc:86}
{EMC Collaboration, M.~Arneodo {\etal}, {\ZF{31}{1986}{1}};
  {\PL{B165}{1985}{222}}}.

\bibitem{jones:91}
{WA21 Collaboration, C.T.~Jones {\etal}, {\ZF{51}{1991}{11}}}.

\bibitem{Dahl89}
{ P.~Dahlqvist, G.~Gustafson and B.~Andersson, {\NP{B328}{1989}{76}};\\ G.
  Gustafson and A. Nilsson, {\NP{B355}{1991}{106}};\\ G. Gustafson and C.
  Sj\"ogren, {\PL{B248}{1990}{430}}}.

\bibitem{lund:anom}
{ B.~Andersson {\etal}, {\ZF{49}{1991}{79}}}.

\bibitem{ringwald}
{A.~Ringwald and F.~Schrempp, DESY-94-197 (1994);\\ M.J.~Gibbs, A.~Ringwald and
  F.~Schrempp, DESY-95-119 (1995)}.

\bibitem{weiner:et}
{M.~Pl\"umer and R.M.~Weiner, {\PRV{D37}{1988}{3136}}}.

\bibitem{na22:neg:binom1}
{NA22 Collaboration, M.~Adamus {\it et al.\/}, {\ZF{37}{1988}{215}}}.

\bibitem{adr:thesis}
{A.~De~Roeck, Ph.D. thesis, Univ. of Antwerpen, 1988 (unpublished)}.

\bibitem{sfm:mul}
{A.~Breakstone {\etal}, {\NC{A102}{1989}{1199}}}.

\bibitem{ua1:mul}
{UA1 Collaboration, C.~Albajar {\etal}, {\NP{B335}{1990}{261}}}.

\bibitem{h1:dis:photo}
{H1 Collaboration, S.~Aid {\etal}, {\PL{B358}{1995}{412}}}.

\bibitem{pythia}
{H.U. Bengtsson and T. Sj\"ostrand, Comp. Phys. Com. {\rm 46} (1987) 43}.

\bibitem{capella:dpm}
{A.~Capella and J.~Tran Thanh Van, {\ZF{23}{1984}{165}}}.

\bibitem{fritiof}
{B.Andersson, G.Gustafson and B.Nilsson-Almqvist, {\NP{B281}{1987}{289}}}.

\bibitem{h1:etflow:95}
{H1 Collaboration, S.~Aid {\etal}, {\PL{B356}{1995}{118}}}.

\end{thebibliography}
